\documentclass[12pt]{article}
\newcommand{\bq}{\begin{equation}}
\newcommand{\eq}{\end{equation}}
\newcommand{\ba}{\begin{eqnarray}}
\newcommand{\ea}{\end{eqnarray}}

\newcommand{\nl }{ \nonumber  }

\newcommand{\p}{\partial}
\newcommand{\h}{\hspace{.5cm}}
\newcommand{\s}{\sigma}

\begin{document}
\vspace*{.5cm}
\begin{center}
{\bf EXACT ROTATING MEMBRANE SOLUTIONS ON A $G_2$ MANIFOLD
AND THEIR SEMICLASSICAL LIMITS
\vspace*{0.5cm}\\ P. Bozhilov}
\\ {\it Institute for Nuclear Research and Nuclear Energy, \\
Bulgarian Academy of Sciences, \\ 1784 Sofia, Bulgaria\\
E-mail:} bozhilov@inrne.bas.bg
\end{center}
\vspace*{0.5cm}

We obtain exact rotating membrane solutions and explicit expressions for the conserved charges on a manifold with exactly known metric of $G_2$ holonomy in M-theory, with four dimensional $\mathcal{N}=1$ gauge theory dual. After that, we investigate their semiclassical limits and derive different relations between the energy and the other conserved quantities, which is a step towards M-theory lift of the semiclassical string/gauge theory correspondence
for $\mathcal{N}=1$ field theories.


\vspace*{.5cm} 

\hspace{2.4cm}$G_2$ manifolds. 


\section{Introduction}
The paper \cite{4} by Gubser, Klebanov and Polyakov on the semiclassical
limit of the string/ gauge theory duality initiated also an interest in the investigation of the $M$-theory lift of this semiclassical correspondence and in particular, in obtaining new membrane solutions in curved space-times and relating their energy and other conserved charges to the dual objects on the field theory side \cite{6}-\cite{0508}. 

M2-brane configurations in $AdS_7\times S^4$ space-time, with field theory dual $A_{N-1}(2,0)$ $SCFT$, have been considered in \cite{6}-\cite{14} and \cite{JHEP032}. In \cite{6}, rotating membrane solution in $AdS_7$ have been obtained. Rotating and boosted membrane configurations was investigated in \cite{10}. Multiwrapped circular membrane, pulsating in the radial direction of $AdS_7$, has been considered in \cite{14}. 
A number of new membrane solutions have been found in \cite{JHEP032} and compared with the already known ones.
Membrane configurations in $AdS_4\times S^7$, $AdS_4\times Q^{1,1,1}$, warped $AdS_5\times M^6$ and in 11-dimensional $AdS$-black hole backgrounds have been considered in \cite{27}.\footnote{See also \cite{MS1} and \cite{MS2}.} 
In \cite{HT04} and \cite{BRR04}, new membrane solutions in $AdS_p\times S^q$ have been also obtained, by using different type of membrane embedding.\footnote{The same type of embedding was previously used in \cite{AFP02} for obtaining new membrane solutions in flat space-time.}
An approach for obtaining exact membrane solutions in general M-theory backgrounds, having field theory dual description has been proposed in \cite{B0705}. As an application, several types of membrane solutions in $AdS_4\times S^7$ background have been found. In a recent paper \cite{0508}, $p$-branes in $AdS_D$ have been examined in two limits, where they exhibit partonic behavior. Namely, rotating branes with energy concentrated to cusp-like solitons and tensionless branes with energy distributed over singletonic bits on the Dirac hypercone. Evidence for a smooth transition from cusps to bits have been found.

To our knowledge, the only paper devoted to rotating membranes on $G_2$ manifolds is \cite{27}, where various membrane configurations on different $G_2$ holonomy backgrounds have been studied systematically. In the semiclassical limit (large conserved charges), the following relations between the energy and the corresponding charge $K$ have been obtained: $E\sim K^{1/2}$, $E\sim K^{2/3}$, $E-K\sim K^{1/3}$, $E-K\sim \ln K$.

Here, our approach will be different. Taking into account that only a small number of $G_2$ holonomy metrics are known {\it exactly}, we choose to search for rotating membrane solutions on one of these metrics, namely, the one discovered in \cite{NPB611}. In section 2, we describe the $G_2$ holonomy background of \cite{NPB611} and its reduction to type IIA string theory. In section 3, we settle the framework, which we will work in. In section 4, we obtain  a number of exact rotating membrane solutions and the explicit expressions for the corresponding conserved charges. Then, we take the semiclassical limit and derive different energy-charge relations. They reproduce and generalize part of the results obtained in \cite{27}, for the case of more than one conserved charges. Section 5 is devoted to our concluding remarks.

\setcounter{equation}{0}
\section{The $G_2$ holonomy background and its type IIA reduction}
The background is a one-parameter family of $G_2$ holonomy metrics (parameterized by $r_0$), which play an important role as supergravity dual of the large $N$ limit of four dimensional $\mathcal{N}=1$ supersymmetric Yang-Mills. These metrics describe the M~theory lift of the supergravity solution corresponding to a collection of D6-branes wrapping the supersymmetric three-cycle of the deformed conifold geometry for any value of the string coupling constant. The explicit expression for the metric with $SU(2)\times SU(2)\times U(1)\times Z_2$ symmetry is given by \cite{NPB611}
\ba\label{G2-1}
ds^2_7 = \sum_{a=1}^7 e^a\otimes e^a,\ea
with the following vielbeins
\ba\nl
e^1 & = &A(r) (\sigma_1-\Sigma_1) ~,~~e^2 = A(r) (\sigma_2-\Sigma_2) ~, \\ \nl
e^3 & = &D(r) (\sigma_3-\Sigma_3) ~,~~e^4 = B(r) (\sigma_1+\Sigma_1) ~, \\ \nl
e^5 & = &B(r) (\sigma_2+\Sigma_2) ~,~~e^6 = r_0 C(r)(\sigma_3+\Sigma_3) ~, \\ \label{G2-2}
e^7 & = &dr/C(r),\ea
where 
\ba\nl
A&=&\frac{1}{\sqrt{12}} \sqrt{(r - 3 r_0/2)(r + 9 r_0/2)},\h
B=\frac{1}{\sqrt{12}} \sqrt{(r + 3 r_0/2)(r - 9 r_0/2)},
\\ \label{G2-3} C&=&\sqrt{\frac{(r - 9 r_0/2)(r + 9 r_0/2)}{(r - 3 r_0/2)(r + 3 r_0/2)}},\h D=r/3,\ea
and 
\ba\nl &&\sigma_1 = \sin \psi \sin \theta d \phi+\cos \psi d \theta,
\h\Sigma_1 = \sin \tilde \psi \sin \tilde \theta d \tilde \phi+\cos \tilde \psi d \tilde \theta
, \\ \nl
&&\sigma_2 =\cos \psi \sin \theta d \phi - \sin \psi d \theta,
\h \Sigma_2 = \cos \tilde \psi \sin \tilde \theta d \tilde \phi- \sin \tilde \psi d \tilde \theta
, \\ \label{G2-4}
&&\sigma_3 = \cos \theta d \phi+d \psi,\hspace{2.2cm}
\Sigma_3 = \cos \tilde \theta d \tilde \phi+d \tilde \psi.\ea
This metric is Ricci flat and complete for $r \geq 9r_0/2$. 
It has a $G_2$-structure given by the following covariantly constant three-form
\ba\nl
\Phi&=&{9r_0^3\over 16}\epsilon_{abc}\;\left(\sigma_a\wedge\sigma_b\wedge\sigma_c-
\Sigma_a\wedge\Sigma_b\wedge\Sigma_c\right)
\\ \nl &+& d\left[{r\over 18}\left(r^2-{27r_0^2\over 4}\right)\left(\sigma_1\wedge \Sigma_1+ \sigma_2\wedge
\Sigma_2\right)+ {r_0\over 3}\left(r^2-{81r_0^2\over 8}\right)\sigma_3\wedge \Sigma_3\right],\ea
which guarantees the existence of a unique covariantly constant spinor \cite{NPB611}.

The metric under consideration is a $U(1)$ bundle over
a six-dimensional manifold. The circle, parameterized by the vielbein $e^6$, 
has its size at infinity set by $r_0$, because $C\rightarrow 1$ when $r\rightarrow \infty$. 
Let us note that the size of the circle at infinity, determines the
Type IIA string coupling constant \cite{NPB611}.
For $r\rightarrow 9r_0/2$, $C\rightarrow 0$ and the circle shrinks to zero size. 

In order to obtain the behavior of the metric for $r\rightarrow \infty$ and $r\rightarrow 9r_0/2$, 
one can rewrite it as follows
\ba\label{G2-g} ds^2_7=dr^2/C^2+A^2((g^1)^2+(g^2)^2)+B^2((g^3)^2+(g^4)^2)+D^2(g^5)^2+r_0\ C^2 (g^6)^2,\ea
where 
\ba\nl 
&&g^1=-\sin\theta_1d\phi_1-\cos\psi_1\sin\theta_2d\phi_2+\sin\psi_1 d\theta_2,\\ \nl
&&g^2=d\theta_1-\sin\psi_1\sin\theta_2 d\phi_2-\cos\psi_1 d\theta_2,\\ \nl
&&g^3=-\sin\theta_1d\phi_1+\cos\psi_1\sin\theta_2d\phi_2-\sin\psi_1 d\theta_2,\\ \nl
&&g^4=d\theta_1+\sin\psi_1\sin\theta_2 d\phi_2+\cos\psi_1 d\theta_2,\\ \nl
&&g^5=d\psi_1+\cos\theta_1 d\phi_1+\cos\theta_2 d\phi_2,\\ \nl
&&g^6=d\psi_2+\cos \theta_1 d\phi_1-\cos\theta_2 d\phi_2.\ea
Then the asymptotic behavior of the metric at infinity is given by \cite{NPB611}
\ba\nl
ds^2=dr^2+r^2\left[{1\over 9}\left(d\psi_1+\sum_{i=1}^{2}\cos \theta_i
d\phi_i\right)^2+{1\over 6}\sum_{i=1}^2\left(d\theta_i^2+\sin^2 \theta_i
d\phi_i^2\right)\right]+ r_0(g^6)^2.\ea
This geometry is that of a $U(1)$ bundle over the singular conifold
metric with $SU(3)$ holonomy. The base of the cone is described by the
Einstein metric on the homogeneous space $T^{1,1}=(SU(2)\times
SU(2))/U(1)$ where the $U(1)$ is diagonally embedded along the Cartan
generator of the $SU(2)$'s. Therefore, at infinity the metric is
topologically ${\bf R}_+\times {\bf S}^1\times {\bf S}^2\times {\bf S}^3$.

In the interior, the metric is non-singular everywhere and near $r = 9r_0/2$ it behaves as
\ba\nl
ds^2\sim d\rho^2+{9\over 4}r_0^2\left[(g^1)^2+(g^2)^2+(g^5)^2\right]+
{\rho^2\over 16}\left[(g^3)^2+(g^4)^2+(g^6)^2\right],\ea
where $\rho^2=8r_0(r-9r_0/2)$. Hence, there exist an ${\bf S}^3$ of finite
size and topologically the space becomes ${\bf R}^4 \times {\bf S}^3$.
As far as $A=D$ and $B=C$ as $r\rightarrow 9r_0/2$, in the interior the metric has enhanced 
$SU(2)\times SU(2)\times SU(2)\times Z_2$ symmetry.
It can be shown \cite{NPB611}, that the metric we get when $r\rightarrow 9r_0/2$, is
the previously known asymptotically conical metric of $G_2$ holonomy on the spin bundle
over ${\bf S}^3$ \cite{BS}-\cite{CGLP}. 

An interesting particular case is when the function $C$ in (\ref{G2-g}) vanishes, 
and the metric of the resulting six-dimensional manifold is given by \cite{NPB611}
\ba\label{6} ds^2_6&=&dt^2 + A^2\left[(\sigma_1-\Sigma_1)^2+(\sigma_2-\Sigma_2)^2\right]+
B^2\left[(\sigma_1+\Sigma_1)^2+(\sigma_2+\Sigma_2)^2\right]
\\ \nl &+&D^2(\sigma_3-\Sigma_3)^2,\h dr=Cdt.\ea 
We note that setting $C=0$ reduces the symmetry 
to $SU(2)\times SU(2)\times Z_2$ which is precisely the symmetry
of the deformed conifold. In this way, one recovers the known metric of $SU(3)$ holonomy 
on the deformed conifold geometry \cite{CO}. Actually, after appropriate change of the coordinates \cite{NPB611},
the metric (\ref{6}) takes the form \cite{KS}
\ba\nl
ds^2_6=K(\tau)\left\{{1\over 3 K^3(\tau)}\left[d\tau^2+(g^5)^2\right]+{1\over 4}\sinh^2\left({\tau/2}\right)[(g^1)^2+(g^2)^2]\right. 
\\ \nl + \left.{1\over 4}\cosh^2\left({\tau/ 2}\right)[(g^3)^2+(g^4)^2]\right\},\ea
where
\ba\nl
K(\tau)={\left[\sinh(2\tau)/2-\tau\right]^{1/3}\over \sinh(\tau)}.\ea
Asymptotically, this metric is also conical and the base of the cone is
topologically ${\bf S}^2\times {\bf S}^3$.

The metric (\ref{G2-1})-(\ref{G2-4}) can be used to describe a four-dimensional
vacuum of the type ${\bf R}^{1,3}\times {\bf X}_7$, where ${\bf X}_7$ is the $G_2$ manifold, 
with four-dimensional ${\cal N}=1$ supersymmetry. The metric under consideration has a $U(1)$
isometry which acts by shifts on an angular coordinate. Hence, one
can reduce it along this $U(1)$ isometry to obtain a Type IIA solution by using that
\ba\nl
ds_{11}^2=e^{-2\phi/3}ds_{10}^2+e^{4\phi/3}(dx_{11}+C_\mu dx^\mu)^2,\ea
where $\phi$ and $C_\mu$ are the Type IIA dilaton and
Ramond-Ramond one-form gauge field respectively. If we identify $x_{11}$ with $\psi_2$,
the reduction to ten dimensions give the following Type IIA solution \cite{NPB611}
\ba\nl
&&ds_{10}^2 = r_0^{1/2}C \left\{dx^2_{1,3} + A^2 \left[ (g^1)^2 +
(g^2)^2 \right] + B^2 \left[ (g^3)^2 + (g^4)^2 \right] + D^2 (g^5)^2 \right\} +
r_0^{1/2}\frac{dr^2}{C}, 
\\ \label{10db} &&e^\phi = r_0^{3/4}C^{3 \over 2},\h
F_2 = \sin \theta_1 d\phi_1 \wedge d\theta_1 -
\sin \theta_2 d\phi_2 \wedge d\theta_2 .\ea
This solution describes a D6-brane wrapping the ${\bf S}^3$ in the
deformed conifold geometry. For $r\to\infty$, the Type IIA metric becomes
that of a singular conifold, the dilaton is constant, and the flux is through the ${\bf S}^2$
surrounding the wrapped D6-brane. 
For $r - 9r_0/2 = \epsilon \to 0$, the string coupling $e^\phi$ goes to zero 
like $\epsilon^{3 \over 4}$, whereas the curvature blows up as
$\epsilon^{- {3 \over 2}}$ just like in the near horizon region
of a flat D6-brane. This means that classical supergravity is valid for sufficiently
large radius. However, the singularity in the interior is the same as the
one of flat D6 branes, as expected. On the other hand, the dilaton continuously decreases 
from a finite value at infinity to zero, 
so that for small $r_0$ classical string theory is valid everywhere.
As explained in \cite{NPB611}, the global geometry is that of a warped product of flat Minkowski
space and a non-compact space, $Y_6$, which for large radius is simply the conifold since the
backreaction of the wrapped D6 brane becomes less and less important. 
However, in the interior, the  backreaction induces changes on $Y_6$ away from the conifold geometry.
For $r \to 9 r_0/2$, the  ${\bf S}^2$ shrinks to zero size, whereas an ${\bf S}^3$ of finite size remains. 
This behavior is similar to that of the deformed conifold but the two metrics are different.
If one mod out the initial eleven-dimensional metric  by
the following $Z_N$ action \cite{NPB611}
\ba\nl Z_N \colon \psi_2 \to \psi_2+ \pi/N\ea
with fixed points located on the ${\bf S}^3$,
then the size of the circle parameterized by $\psi_2$ goes to zero.
As a result, the local geometry at $r \approx 9 r_0/2$ becomes singular,
with $A_{N-1}$ singularity fibered over ${\bf S}^3$, i.e.
the so-called singular quotient \cite{BA}, \cite{AMV}. 
After compactification to Type IIA theory, it describes $N$ coincident
D6-branes wrapped on the supersymmetric ${\bf S}^3$  of the deformed conifold.

\setcounter{equation}{0}
\section{The approach}
In this section, we settle the framework, which we will work in. Actually, we will use 
the general approach developed in \cite{B0705}.

We start with the following membrane action
\ba\label{oma} S= \int d^{3}\xi\mathcal{L} = 
\int d^{3}\xi\left\{\frac{1}{4\lambda^0}\Bigl[G_{00}-2\lambda^{j}G_{0j}+\lambda^{i}
\lambda^{j}G_{ij}-\left(2\lambda^0T_2\right)^2\det G_{ij}\Bigr] + 
T_2 B_{012}\right\},\ea 
where \ba\nl G_{mn}= g_{MN}(X)\p_m X^M\p_n X^N,\h
B_{012}= b_{MNP}(X)\p_{0}X^{M}\p_{1}X^{N}\p_{2}X^{P}, \\ \nl 
\p_m=\p/\p\xi^m,\h m = (0,i) = (0,1,2),\h M = (0,1,\ldots,10),\ea 
are the fields induced on the membrane worldvolume,
$\lambda^m$ are Lagrange multipliers, $x^M=X^M(\xi)$ are the membrane
embedding coordinates, and $T_2$ is its tension. 
As shown in \cite{NPB656}, the above action is
classically equivalent to the Nambu-Goto type action 
\ba\nl S^{NG}= - T_2\int d^{3}\xi
\left(\sqrt{-\det{G_{mn}}}-\frac{1}{6}\varepsilon^{mnp}
\p_{m}X^{M}\p_n X^N \p_{p}X^{P} b_{MNP}\right)\ea
and to the Polyakov type action
\ba\nl S^{P}= - \frac{T_2}{2}\int
d^{3}\xi\left[\sqrt{-\gamma}\left(\gamma^{mn} G_{mn}-1\right) - 
\frac{1}{3} \varepsilon^{mnp}\p_{m}X^{M}\p_nX^N\p_{p}X^{P}b_{MNP}\right],\ea
where $\gamma^{mn}$ is the auxiliary worldvolume metric and $\gamma=\det\gamma_{mn}$.
In addition, the action (\ref{oma}) gives a unified description for the tensile and tensionless membranes.

The equations of motion for the Lagrange multipliers $\lambda^{m}$ generate the constraints
\ba\label{00} &&G_{00}-2\lambda^{j}G_{0j}+\lambda^{i}\lambda^{j}G_{ij}
+\left(2\lambda^0T_2\right)^2\det G_{ij}=0,\\
\label{0j} &&G_{0j}-\lambda^{i}G_{ij}=0.\ea
Further on, we will work in the worldvolume gauge $\lambda^{i}=0$, $\lambda^{0}=const$ in which the action (\ref{oma}) and the constraints (\ref{00}), (\ref{0j}) simplify to 
\ba\label{omagf} &&S_{gf}= \int d^{3}\xi\left\{\frac{1}{4\lambda^0}\Bigl[G_{00}-\left(2\lambda^0T_2\right)^2\det G_{ij}\Bigr] + T_2 B_{012}\right\},
\\ \label{00gf} &&G_{00}+\left(2\lambda^0T_2\right)^2\det G_{ij}=0,
\\ \label{0igf} &&G_{0i}=0.\ea
Let us note that the action (\ref{omagf}) and the constraints (\ref{00gf}), (\ref{0igf}) {\it coincide} with the usually used gauge fixed Polyakov type action and constraints after the following identification of the parameters (see for instance \cite{27})
\ba\nl 2\lambda^0T_2=L.\ea

Supposing that there exist a (non-fixed) number of commuting Killing vectors $\p/\p x^\mu$, which leads to
\ba\label{ob} \p_\mu g_{MN} =0,\h \p_\mu b_{MNP} =0,\ea
we will search for {\it rotating} membrane solutions in the framework of the following embedding ($X^M=(X^\mu,X^a)$, $\Lambda^\mu_m = constants$)
\ba\label{sLA} X^\mu(\xi^m)=X^\mu(\tau, \delta, \sigma)=\Lambda^\mu_m \xi^m=
\Lambda^\mu_0\tau+\Lambda^\mu_1\delta+ \Lambda^\mu_2\s,\h
X^a(\xi^m)=Z^a(\s).\ea 
The above ansatz reduces the Lagrangian density in the action (\ref{omagf}) to $(Z'^a=dZ^a/d\s)$
\ba\label{old} \mathcal{L}^{A}(\sigma) =\frac{1}{4\lambda^0}\left[K_{ab}(g)Z'^aZ'^b + 
2A_{a}(g,b)Z'^a - V(g,b)\right],\ea where 
\ba\nl &&K_{ab}(g)=-\left(2\lambda^0T_2\right)^2 \Lambda^\mu_1\Lambda^\nu_1
\left(g_{ab}g_{\mu\nu}-g_{a\mu}g_{b\nu}\right),\\ \nl 
&&A_{a}(g,b)=\left(2\lambda^0T_2\right)^2 \Lambda^\mu_1\Lambda^\nu_1\Lambda^\rho_2
\left(g_{a\mu}g_{\nu\rho}-g_{a\rho}g_{\mu\nu}\right)+2\lambda^0T_2\Lambda^\mu_0\Lambda^\nu_1 b_{a\mu\nu},\\ \nl
&&V(g,b)=-\Lambda^\mu_0\Lambda^\nu_0 g_{\mu\nu}+ \left(2\lambda^0T_2\right)^2 \Lambda^\mu_1\Lambda^\nu_1\Lambda^\rho_2\Lambda^\lambda_2
\left(g_{\mu\nu}g_{\rho\lambda}-g_{\mu\rho}g_{\nu\lambda}\right)
-4\lambda^0T_2\Lambda^\mu_0\Lambda^\nu_1\Lambda^\rho_2 b_{\mu\nu\rho}.\ea
$\mathcal{L}^{A}$ does not depend on $\tau$ and $\delta$ because of (\ref{ob}) and (\ref{sLA}).

Now, the constraints (\ref{00gf}) and (\ref{0igf}) can be written in the form
\ba\label{00e} &&K_{ab}Z'^aZ'^b + U=0,
\\ \label{01} &&\Lambda^\mu_0\Lambda^\nu_1 g_{\mu\nu}=0,
\\ \label{02} &&\Lambda^\mu_0\left(g_{\mu a}Z'^a + \Lambda^\nu_2 g_{\mu\nu}\right)=0,\ea
where $U=V + 4\lambda^0\Lambda^\mu_2\mathcal{P}^2_\mu$, and 
\ba\nl 2\lambda^0\mathcal{P}^2_\mu &=& \left(2\lambda^0T_2\right)^2 \Lambda^\nu_1\Lambda^\rho_1\left(g_{\mu\nu}g_{\rho a}-g_{\nu\rho}g_{\mu a}\right)Z'^a
\\ \label{cm} &+&\left(2\lambda^0T_2\right)^2 \Lambda^\nu_1\Lambda^\rho_1\Lambda^\lambda_2
\left(g_{\mu\nu}g_{\rho\lambda}-g_{\mu\lambda}g_{\nu\rho}\right) 
+ 2\lambda^0T_2\Lambda^\nu_0\Lambda^\rho_1 b_{\mu\nu\rho}\ea
are constants of the motion \cite{B0705}.

Due to the independence of $\mathcal{L}^{A}(\sigma)$ on $X^\mu$, the momenta
\ba\label{cmom} P_\mu = \int d^{2}\xi p_\mu 
=\frac{1}{2\lambda^{0}}\int\int d\delta d\sigma\left[\Lambda^\nu_0 g_{\mu\nu} + 2\lambda^0T_2\Lambda^\nu_1\left(b_{\mu\nu a}Z'^a+\Lambda^\rho_2 b_{\mu\nu\rho}\right)\right]\ea
are conserved, i.e. they do not depend on the proper time $\tau$.

In this article, we are interested in obtaining membrane solutions for which the conditions (\ref{01}), (\ref{02}) and $\mathcal{P}^2_\mu=constants$ are satisfied identically by an appropriate choice of the embedding parameters $\Lambda^\mu_m$.Then, the investigation of the membrane dynamics reduces to the problem of solving the equations of motion following from (\ref{old}), which are 
\ba\label{ema} K_{ab}Z''^b + \Gamma^{K}_{a,bc}Z'^b Z'^c - 2\p_{[a}A_{b]}Z'^b  
+ \frac{1}{2}\p_a U = 0,\ea where 
\ba\nl \Gamma^{K}_{a,bc}= \frac{1}{2}\left(\p_b K_{ca}+\p_c K_{ba}-\p_a K_{bc}\right),
\h\p_{[a}A_{b]}=\frac{1}{2}\left(\p_a A_b - \p_b A_a\right), \ea
and the remaining constraint (\ref{00e}). Finally, let us note that if the embedding is such that the background seen by the membrane depends on only one coordinate $x^a$ , then the constraint (\ref{00e}) is first integral of the equation of motion (\ref{ema}) for $X^a(\xi^m)=Z^a(\s)$, and the general solution is given by \cite{B0705}
\ba\label{1dc}\sigma\left(X^a\right)=\sigma_0 + \int_{X_0^a}^{X^a}
\left(-\frac{K_{aa}}{U}\right)^{1/2}dx,\ea 
where $\sigma_0$ and $X_0^a$ are arbitrary constants. Namely this solution will be used in the next section
in the following form
\ba\label{gf} \sigma\left(X^a\right)=\int_{X^a_{min}}^{X^a}
\left(-\frac{K_{aa}}{U}\right)^{1/2}dx.\ea 
Also, the normalization condition
\ba\label{nc} 2\pi=\int_0^{2\pi}d\s=2\int_{X^a_{min}}^{X^a_{max}}\left(-\frac{K_{aa}}{U}\right)^{1/2}dx\ea 
will be imposed, which means that the two periods must be equal.

\setcounter{equation}{0}
\section{Exact rotating membrane solutions and their semiclassical limits}
The M-theory background, which we will use from now on, has the form
\ba\label{11db} l_{11}^{-2}ds_{11}^{2}=-dt^2 + \delta_{IJ}dx^I dx^J + ds_{7}^{2},\ea 
where $l_{11}$ is the eleven dimensional Planck length, \small{({\it I,J=1,2,3})} and $ds_{7}^{2}$ is given in (\ref{G2-1})-(\ref{G2-4}). In other words, the background is direct product of flat, four dimensional space-time, and a seven dimensional $G_2$ manifold. 

As already mentioned above, we will search for solutions, for which the background felt by the membrane depends on only one coordinate. This will be the radial coordinate $r$, i.e. the rotating membrane embedding along this coordinate has the form $r=r(\s)$. Then, according to our ansatz (\ref{sLA}), the remaining membrane coordinates, which are not fixed, will depend linearly on the worldvolume coordinates $\tau$, $\delta$ and $\sigma$. The membrane configurations considered below are all for which, we were able to obtain {\it exact} solutions under the described conditions.

\subsection{First type of membrane embedding}
Let us consider the following membrane configuration:
\ba\nl
&&X^0\equiv t=\Lambda_0^0\tau+\frac{1}{\Lambda_0^0}\left[\left(\mathbf{\Lambda}_0.\mathbf{\Lambda}_1\right)\delta + \left(\mathbf{\Lambda}_0.\mathbf{\Lambda}_2\right)\sigma\right],
\h X^I=\Lambda_0^I\tau + \Lambda_1^I\delta + \Lambda_2^I\sigma, 
\\ \label{A1} &&X^4\equiv r(\s),\h X^6\equiv \theta=\Lambda_0^6\tau,\h X^9\equiv\tilde{\theta}=\Lambda_0^9\tau;
\h \left(\mathbf{\Lambda}_0.\mathbf{\Lambda}_i\right)=\delta_{IJ}\Lambda_0^I\Lambda_i^J.\ea
It corresponds to membrane extended in the radial direction $r$, and rotating in the planes given by the angles  $\theta$ and $\tilde{\theta}$. In addition, it is nontrivially spanned along $X^0$ and $X^I$. The relations between the parameters in $X^0$ and $X^I$ guarantee that the equalities (\ref{01}), (\ref{02}) and $\mathcal{P}^2_\mu=constants$ are identically satisfied.
At the same time, the membrane moves along $t$-coordinate with constant energy $E$, and along $X^I$ with constant momenta $P_{I}$. In this case, the target space metric seen by the membrane becomes 
\ba\nl &&g_{00}\equiv g_{tt}=-l_{11}^{2},\h g_{IJ}=l_{11}^{2}\delta_{IJ},
\h g_{44}\equiv g_{rr}=\frac{l_{11}^{2}}{C^2(r)},
\\ \nl &&g_{66}\equiv g_{\theta\theta}=l_{11}^{2}\left[A^2(r)+B^2(r)\right],\h
g_{99}\equiv g_{\tilde{\theta}\tilde{\theta}}=l_{11}^{2}\left[A^2(r)+B^2(r)\right],\\ \label{b1}
&&g_{69}\equiv g_{\theta\tilde{\theta}}=-l_{11}^{2}\left[A^2(r)-B^2(r)\right].\ea
Therefore, in the notations introduced in (\ref{sLA}), we have 
$\mu=(0,I,6,9)\equiv(t,I,\theta,\tilde{\theta})$, $a=4\equiv r$.
The metric induced on the membrane worldvolume is
\ba\nl &&G_{00}=-l_{11}^{2}\left[(\Lambda_0^0)^2-\mathbf{\Lambda}_0^2 
- (\Lambda_0^-)^2 A^2-(\Lambda_0^+)^2 B^2\right],\\ \nl 
&&G_{11}=l_{11}^{2}M_{11},\h G_{12}=l_{11}^{2}M_{12},\h G_{22}=l_{11}^{2}\left[M_{22} + \frac{r'^2}{C^2}\right],\ea
where 
\ba \label{DM} M_{ij}= \left(\mathbf{\Lambda}_i.\mathbf{\Lambda}_j\right)
-\frac{\left(\mathbf{\Lambda}_0.\mathbf{\Lambda}_i\right)
\left(\mathbf{\Lambda}_0.\mathbf{\Lambda}_j\right)}{\left(\Lambda_0^0\right)^2},
\h \Lambda_0^{\pm}=\Lambda_0^6\pm\Lambda_0^9.\ea
The constants of the motion $\mathcal{P}^2_\mu$, introduced in (\ref{cm}), are given by
\ba\label{cm1} &&\mathcal{P}^2_0=-\frac{2\lambda^0 T_2^2 l_{11}^4}{\Lambda_0^0}
\left[\left(\mathbf{\Lambda}_0.\mathbf{\Lambda}_1\right)M_{12} - \left(\mathbf{\Lambda}_0.\mathbf{\Lambda}_2\right)M_{11}\right],\\ \nl
&&\mathcal{P}^2_I=2\lambda^0 T_2^2 l_{11}^4\left(\Lambda_1^I M_{12}-\Lambda_2^I M_{11}\right),\h
\mathcal{P}^2_6=\mathcal{P}^2_9=0.\ea
The Lagrangian (\ref{old}) takes the form
\ba\nl &&\mathcal{L}^{A}(\sigma) =\frac{1}{4\lambda^0}\left(K_{rr}r'^2 - V\right),\h
K_{rr}=-(2\lambda^0 T_2 l_{11}^2)^2\frac{M_{11}}{C^2},\\ \nl
&&V=(2\lambda^0 T_2 l_{11}^2)^2 \det M_{ij} + 
l_{11}^{2}\left[(\Lambda_0^0)^2-\mathbf{\Lambda}_0^2 
- (\Lambda_0^-)^2 A^2-(\Lambda_0^+)^2 B^2\right].\ea

Let us first consider the particular case when $\Lambda_0^-=0$, i.e. $\theta=\tilde{\theta}$. 
From the yet unsolved constraint (\ref{00e})
\ba\nl K_{rr}r'^2 + U=0,\h U=V + 4\lambda^0\Lambda^\mu_2\mathcal{P}^2_\mu ,\ea
one obtains the turning points of the effective one-dimensional periodic motion by solving the equation $r'=0$.
In the case under consideration, the result is
\ba\nl &&r_{min}=3l,\h  
r_{max}=r_1=l\left(2\sqrt{1+\frac{3u_0^2}{l^2(\Lambda_0^+)^2}}+1\right)>3l,
\\ \nl &&r_2=-l\left(2\sqrt{1+\frac{3u_0^2}{l^2(\Lambda_0^+)^2}}-1\right)<0, \h l=3r_0/2,\ea
where we have introduced the notation
\ba\label{u02} u_0^2&=&(2\lambda^0 T_2 l_{11})^2\det M_{ij} + 
(\Lambda_0^0)^2-\mathbf{\Lambda}_0^2 + 4\lambda^0\Lambda^\mu_2\mathcal{P}^2_\mu/l_{11}^2
\\ \nl &=&(\Lambda_0^0)^2-\mathbf{\Lambda}_0^2 - (2\lambda^0 T_2 l_{11})^2\det M_{ij}. \ea 

Applying the general formula (\ref{gf}), we obtain the following expression for the membrane solution 
($\Delta r=r-3l$)
\ba\nl &&\sigma(r)=\int_{3l}^{r}\left[-\frac{K_{rr}(t)}{U(t)}\right]^{1/2}dt= 
\frac{16\lambda^0 T_2 l_{11}}{\Lambda_0^+}\left[\frac{M_{11}l\Delta r}{\left(r_1-3l\right)\left(3l-r_2\right)}\right]^{1/2}\times
\\ \label{s1} &&F_D^{(5)}\left(1/2;-1/2,-1/2,1/2,1/2,1/2;3/2;-\frac{\Delta r}{2l},-\frac{\Delta r}{4l},-\frac{\Delta r}{6l},-\frac{\Delta r}{3l-r_2},\frac{\Delta r}{r_1-3l}\right),\ea
where $F_D^{(5)}$ is a hypergeometric function of five variables. The definition and some properties of the hypergeometric functions $F_D^{(n)}(a;b_1,\ldots,b_n;c;z_1,\ldots,z_n)$ are given in Appendix A.

The normalization condition (\ref{nc}) leads to ($\Delta r_1=r_1-3l$)
\ba\nl &&2\pi=2\int_{3l}^{r_1}\left[-\frac{K_{rr}(t)}{U(t)}\right]^{1/2}dt= 
\frac{32\lambda^0 T_2 l_{11}\left(M_{11}l\right)^{1/2}}{\Lambda_0^+\left(3l-r_2\right)^{1/2}}\times
\\ \nl &&F_D^{(5)}\left(1/2;-1/2,-1/2,1/2,1/2,1/2;3/2;-\frac{\Delta r_1}{2l},-\frac{\Delta r_1}{4l},-\frac{\Delta r_1}{6l},-\frac{\Delta r_1}{3l-r_2},1\right)=
\\ \nl &&\frac{16\pi\lambda^0 T_2 l_{11}\left(M_{11}l\right)^{1/2}}{\Lambda_0^+\left(3l-r_2\right)^{1/2}}
F_D^{(4)}\left(1/2;-1/2,-1/2,1/2,1/2,;1;-\frac{\Delta r_1}{2l},-\frac{\Delta r_1}{4l},-\frac{\Delta r_1}{6l},-\frac{\Delta r_1}{3l-r_2}\right)
\\ \nl &&=\frac{16\pi\lambda^0 T_2 l_{11}\left(M_{11}l\right)^{1/2}}{\Lambda_0^+\left(3l-r_2\right)^{1/2}}
\left(1+\frac{\Delta r_1}{2l}\right)^{1/2}\left(1+\frac{\Delta r_1}{4l}\right)^{1/2} \left(1+\frac{\Delta r_1}{6l}\right)^{-1/2}\left(1+\frac{\Delta r_1}{3l-r_2}\right)^{-1/2}
\\ \label{nc1} &&\times F_D^{(4)}\left(1/2;-1/2,-1/2,1/2,1/2,;1;\frac{1}{1+\frac{2l}{\Delta r_1}}, \frac{1}{1+\frac{4l}{\Delta r_1}}, \frac{1}{1+\frac{6l}{\Delta r_1}},\frac{1}{1+\frac{3l-r_2}{\Delta r_1}}\right).\ea

Now, we can compute the conserved momenta on the obtained solution. According to (\ref{cmom}), they are:
\ba\label{EP1} &&E=-P_0=\frac{\pi^2l_{11}^2}{\lambda^0}\Lambda_0^0,\h \mathbf{P}=\frac{\pi^2l_{11}^2}{\lambda^0}\mathbf{\Lambda}_0,
\\ \nl &&P_{\theta}=P_{\tilde{\theta}}=\frac{\pi l_{11}^2}{\lambda^0}\Lambda_0^+ \int_{3l}^{r_1}\left[-\frac{K_{rr}(t)}{U(t)}\right]^{1/2}B^2(t)dt= 
\frac{4\pi^2T_2 l_{11}^3\left(M_{11}l^3\right)^{1/2}}{3\left(3l-r_2\right)^{1/2}}\times
\\ \nl &&\Delta r_1 F_D^{(4)}\left(3/2;-1/2,-3/2,1/2,1/2,;2;-\frac{\Delta r_1}{2l},-\frac{\Delta r_1}{4l},-\frac{\Delta r_1}{6l},-\frac{\Delta r_1}{3l-r_2}\right)
\\ \nl &&=\frac{4\pi^2T_2 l_{11}^3\left(M_{11}l^3\right)^{1/2}}{3\left(3l-r_2\right)^{1/2}}\Delta r_1
\left(1+\frac{\Delta r_1}{2l}\right)^{1/2}\left(1+\frac{\Delta r_1}{4l}\right)^{3/2} \left(1+\frac{\Delta r_1}{6l}\right)^{-1/2}\left(1+\frac{\Delta r_1}{3l-r_2}\right)^{-1/2}
\\ \label{cmom1} &&\times F_D^{(4)}\left(1/2;-1/2,-3/2,1/2,1/2,;2;\frac{1}{1+\frac{2l}{\Delta r_1}}, \frac{1}{1+\frac{4l}{\Delta r_1}}, \frac{1}{1+\frac{6l}{\Delta r_1}},
\frac{1}{1+\frac{3l-r_2}{\Delta r_1}}\right).\ea

Our next task is to find the relation between the energy $E$ and the other conserved quantities $\mathbf{P}$, $P_{\theta}=P_{\tilde{\theta}}$ in the semiclassical limit (large conserved charges). This corresponds to $r_1\to\infty$, which in the present case leads to $3u_0^2/[l^2(\Lambda_0^+)^2]\to\infty$. In this limit, the condition (\ref{nc1}) reduces to
\ba\nl \Lambda_0^+ = 2\sqrt{3}\lambda^0 T_2 l_{11}M_{11}^{1/2},\ea
while the expression (\ref{cmom1}) for the momentum $P_{\theta}$, takes the form
\ba\nl P_{\theta}=P_{\tilde{\theta}}= \sqrt{3}\pi^2 T_2 l_{11}^3 M_{11}^{1/2}\frac{u_0^2}{(\Lambda_0^+)^2}.\ea
Combining these results with (\ref{EP1}), one obtains 
\ba\label{Eg1} &&\left\{E^2\left(E^2-\mathbf{P}^2\right) - (2\pi^2 T_2 l_{11}^3)^2\left\{\left(\mathbf{\Lambda}_1\times\mathbf{\Lambda}_2\right)^2 E^2 - \left[\left(\mathbf{\Lambda}_1\times\mathbf{\Lambda}_2\right)\times\mathbf{P}\right]^2\right\}\right\}^2 
\\ \nl &&-(4\sqrt{3}\pi^2 T_2 l_{11}^3)^2E^2\left[\mathbf{\Lambda}_1^2 E^2 - \left(\mathbf{\Lambda}_1.\mathbf{P}\right)^2\right]P_{\theta}^2 = 0,
\h \left(\mathbf{\Lambda}_1\times\mathbf{\Lambda}_2\right)_I = \varepsilon_{IJK}\Lambda_1^J\Lambda_2^K.\ea
This is {\it fourth} order algebraic equation for $E^2$. Its positive solutions give the explicit dependence of the energy on $\mathbf{P}$ and $P_{\theta}$: $E^2=E^2(\mathbf{P}, P_{\theta})$. 

Let us consider a few particular cases.
In the simplest case, when $\Lambda_0^I=0$, i.e. $\mathbf{P}=0$, and $\Lambda_2^I=c\Lambda_1^I$, which corresponds to the membrane embedding (see (\ref{A1}))
\ba\nl X^0\equiv t=\Lambda_0^0\tau, \h X^I=\Lambda_1^I(\delta + c\sigma),\h X^4\equiv r(\s),
\h X^6\equiv \theta=\Lambda_0^6\tau = X^9\equiv\tilde{\theta}=\Lambda_0^9\tau,\ea
(\ref{Eg1}) simplifies to
\ba\label{p1} E^2=4\sqrt{3}\pi^2 T_2 l_{11}^3\mid\mathbf{\Lambda}_1\mid P_{\theta}.\ea
This is the relation $E\sim K^{1/2}$ obtained for $G_2$-manifolds in \cite{27}.
If we impose only the conditions $\Lambda_0^I=0$, and $\Lambda_i^I$ remain independent, (\ref{Eg1}) gives
\ba\label{p2} E^2 =(2\pi^2 T_2 l_{11}^3)^2\left(\mathbf{\Lambda}_1\times\mathbf{\Lambda}_2\right)^2  + 4\sqrt{3}\pi^2 T_2 l_{11}^3\mid\mathbf{\Lambda}_1\mid P_{\theta}.\ea
Now, let us take $\Lambda_0^I\ne 0$, $\Lambda_2^I=c\Lambda_1^I$. Then, (\ref{Eg1}) reduces to
\ba\nl E^2\left[\left(E^2-\mathbf{P}^2\right)^2 - (4\sqrt{3}\pi^2 T_2 l_{11}^3)^2\mathbf{\Lambda}_1^2P_{\theta}^2\right]
+(4\sqrt{3}\pi^2 T_2 l_{11}^3)^2 \left(\mathbf{\Lambda}_1.\mathbf{P}\right)^2P_{\theta}^2=0 ,\ea
which is {\it third} order algebraic equation for $E^2$. If the three-dimensional vectors $\mathbf{\Lambda}_1$ and $\mathbf{P}$ are orthogonal to each other, i.e. $\left(\mathbf{\Lambda}_1.\mathbf{P}\right)=0$, the above relation simplifies to
\ba\label{p3} E^2=\mathbf{P}^2 + 4\sqrt{3}\pi^2 T_2 l_{11}^3\mid\mathbf{\Lambda}_1\mid P_{\theta}.\ea
The obvious conclusion is that in the framework of a given embedding, one can obtain different relations between the energy and the other conserved charges, depending on the choice of the embedding parameters.

Now, we will consider the general case, when $\Lambda_0^-\ne 0$, i.e. $\theta\ne\tilde{\theta}$.
The turning points are given by
\ba\nl &&r_{min}=3l,\h  
r_{max}=r_1=l\left[2\sqrt{\frac{k^2+3}{4}+\frac{3u_0^2}{l^2\left((\Lambda_0^+)^2+(\Lambda_0^-)^2\right)}}+k\right],
\\ \nl &&r_2=-l\left[2\sqrt{\frac{k^2+3}{4}+\frac{3u_0^2}{l^2\left((\Lambda_0^+)^2+(\Lambda_0^-)^2\right)}}-k\right], \h k=\frac{(\Lambda_0^+)^2-(\Lambda_0^-)^2}{(\Lambda_0^+)^2+(\Lambda_0^-)^2}\in [0,1].\ea

According to (\ref{gf}), the solution for $\s(r)$ is
\ba\nl &&\sigma(r)=\int_{3l}^{r}\left[-\frac{K_{rr}(t)}{U(t)}\right]^{1/2}dt= 
\frac{16\lambda^0 T_2 l_{11}}{\left[(\Lambda_0^+)^2+(\Lambda_0^-)^2\right]^{1/2}}\left[\frac{M_{11}l\Delta r}{\left(r_1-3l\right)\left(3l-r_2\right)}\right]^{1/2}\times
\\ \label{sg1} &&F_D^{(5)}\left(1/2;-1/2,-1/2,1/2,1/2,1/2;3/2;-\frac{\Delta r}{2l},-\frac{\Delta r}{4l},-\frac{\Delta r}{6l},-\frac{\Delta r}{3l-r_2},\frac{\Delta r}{r_1-3l}\right).\ea

The normalization condition (\ref{nc}) reads 
\ba\nl &&\frac{8\lambda^0 T_2 l_{11}\left(M_{11}l\right)^{1/2}}{\left[(\Lambda_0^+)^2+(\Lambda_0^-)^2\right]^{1/2}\left(3l-r_2\right)^{1/2}}\times
\\ \nl &&F_D^{(4)}\left(1/2;-1/2,-1/2,1/2,1/2,;1;-\frac{\Delta r_1}{2l},-\frac{\Delta r_1}{4l},-\frac{\Delta r_1}{6l},-\frac{\Delta r_1}{3l-r_2}\right)=
\\ \nl &&\frac{8\lambda^0 T_2 l_{11}\left(M_{11}l\right)^{1/2}}{\left[(\Lambda_0^+)^2+(\Lambda_0^-)^2\right]^{1/2}\left(3l-r_2\right)^{1/2}}\times
\\ \nl &&\left(1+\frac{\Delta r_1}{2l}\right)^{1/2}\left(1+\frac{\Delta r_1}{4l}\right)^{1/2} \left(1+\frac{\Delta r_1}{6l}\right)^{-1/2}\left(1+\frac{\Delta r_1}{3l-r_2}\right)^{-1/2}\times 
\\ \label{ncg1} &&F_D^{(4)}\left(1/2;-1/2,-1/2,1/2,1/2,;1;\frac{1}{1+\frac{2l}{\Delta r_1}}, \frac{1}{1+\frac{4l}{\Delta r_1}}, \frac{1}{1+\frac{6l}{\Delta r_1}},\frac{1}{1+\frac{3l-r_2}{\Delta r_1}}\right)=1.\ea

Computing the conserved momenta in accordance with (\ref{cmom}), one obtains the same expressions for $E$ and $\mathbf{P}$ as in (\ref{EP1})\footnote{Actually, these expressions for $E$ and $\mathbf{P}$ are always valid  for the background we use in this paper.}, and 
\ba\nl &&\frac{1}{2}\left(P_{\theta}+P_{\tilde{\theta}}\right)=
\frac{4\pi^2T_2 l_{11}^3\Lambda_0^+\left(M_{11}l^3\right)^{1/2}}
{3\left[(\Lambda_0^+)^2+(\Lambda_0^-)^2\right]^{1/2}\left(3l-r_2\right)^{1/2}}\times
\\ \nl &&\Delta r_1 F_D^{(4)}\left(3/2;-1/2,-3/2,1/2,1/2,;2;-\frac{\Delta r_1}{2l},-\frac{\Delta r_1}{4l},-\frac{\Delta r_1}{6l},-\frac{\Delta r_1}{3l-r_2}\right)
\\ \nl &&=\frac{4\pi^2T_2 l_{11}^3\Lambda_0^+\left(M_{11}l^3\right)^{1/2}}
{3\left[(\Lambda_0^+)^2+(\Lambda_0^-)^2\right]^{1/2}\left(3l-r_2\right)^{1/2}}\times
\\ \nl &&\Delta r_1 \left(1+\frac{\Delta r_1}{2l}\right)^{1/2}\left(1+\frac{\Delta r_1}{4l}\right)^{3/2} \left(1+\frac{\Delta r_1}{6l}\right)^{-1/2}\left(1+\frac{\Delta r_1}{3l-r_2}\right)^{-1/2}\times 
\\ \label{cmomg1t1} &&F_D^{(4)}\left(1/2;-1/2,-3/2,1/2,1/2,;2;\frac{1}{1+\frac{2l}{\Delta r_1}}, \frac{1}{1+\frac{4l}{\Delta r_1}}, \frac{1}{1+\frac{6l}{\Delta r_1}},
\frac{1}{1+\frac{3l-r_2}{\Delta r_1}}\right),\ea

\ba\nl &&\frac{1}{2}\left(P_{\theta}-P_{\tilde{\theta}}\right)=
\frac{8\pi^2T_2 l_{11}^3\Lambda_0^-\left(M_{11}l^5\right)^{1/2}}
{\left[(\Lambda_0^+)^2+(\Lambda_0^-)^2\right]^{1/2}\left(3l-r_2\right)^{1/2}}\times
\\ \nl &&F_D^{(4)}\left(1/2;-3/2,-1/2,-1/2,1/2,;1;-\frac{\Delta r_1}{2l},-\frac{\Delta r_1}{4l},
-\frac{\Delta r_1}{6l},-\frac{\Delta r_1}{3l-r_2}\right)
\\ \nl &&=\frac{8\pi^2T_2 l_{11}^3\Lambda_0^-\left(M_{11}l^5\right)^{1/2}}
{\left[(\Lambda_0^+)^2+(\Lambda_0^-)^2\right]^{1/2}\left(3l-r_2\right)^{1/2}}\times
\\ \nl &&\left(1+\frac{\Delta r_1}{2l}\right)^{3/2}\left(1+\frac{\Delta r_1}{4l}\right)^{1/2} \left(1+\frac{\Delta r_1}{6l}\right)^{1/2}\left(1+\frac{\Delta r_1}{3l-r_2}\right)^{-1/2}\times 
\\ \label{cmomg1t2} &&F_D^{(4)}\left(1/2;-3/2,-1/2,-1/2,1/2,;1;\frac{1}{1+\frac{2l}{\Delta r_1}}, \frac{1}{1+\frac{4l}{\Delta r_1}}, \frac{1}{1+\frac{6l}{\Delta r_1}},
\frac{1}{1+\frac{3l-r_2}{\Delta r_1}}\right).\ea

Now, we go to the semiclassical limit $r_1\to\infty$. The normalization condition (\ref{ncg1}) gives
\ba\nl \left[(\Lambda_0^+)^2+(\Lambda_0^-)^2\right]^{1/2} =  2\sqrt{3}\lambda^0 T_2 l_{11}M_{11}^{1/2},\ea
whereas (\ref{cmomg1t1}) and (\ref{cmomg1t2}) take the form
\ba\nl \frac{1}{2}\left(P_{\theta}\pm P_{\tilde{\theta}}\right)=
\frac{\sqrt{3}\pi^2T_2 l_{11}^3\Lambda_0^\pm M_{11}^{1/2}u_0^2}
{\left[(\Lambda_0^+)^2+(\Lambda_0^-)^2\right]^{3/2}}.\ea
The above expressions, together with (\ref{EP1}), lead to the following connection between the energy and the conserved momenta
\ba\label{Egg1} &&\left\{E^2\left(E^2-\mathbf{P}^2\right) - (2\pi^2 T_2 l_{11}^3)^2\left\{\left(\mathbf{\Lambda}_1\times\mathbf{\Lambda}_2\right)^2 E^2 - \left[\left(\mathbf{\Lambda}_1\times\mathbf{\Lambda}_2\right)\times\mathbf{P}\right]^2\right\}\right\}^2 
\\ \nl &&-6(2\pi^2 T_2 l_{11}^3)^2E^2\left[\mathbf{\Lambda}_1^2 E^2 - \left(\mathbf{\Lambda}_1.\mathbf{P}\right)^2\right]\left(P^2_{\theta}+P^2_{\tilde{\theta}}\right) = 0.\ea
Obviously, (\ref{Egg1}) is the generalization of (\ref{Eg1}) for the case $P_{\theta}\ne P_{\tilde{\theta}}$ and for 
$P_{\theta}=P_{\tilde{\theta}}$ coincides with it, as it should be. The particular cases (\ref{p1}), (\ref{p2}) and (\ref{p3}) now generalize to
\ba\nl &&E^2=2\sqrt{6}\pi^2 T_2 l_{11}^3\mid\mathbf{\Lambda}_1\mid \left(P^2_{\theta}+P^2_{\tilde{\theta}}\right)^{1/2},\\ \nl
&&E^2 =(2\pi^2 T_2 l_{11}^3)^2\left(\mathbf{\Lambda}_1\times\mathbf{\Lambda}_2\right)^2 + 2\sqrt{6}\pi^2 T_2 l_{11}^3\mid\mathbf{\Lambda}_1\mid \left(P^2_{\theta}+P^2_{\tilde{\theta}}\right)^{1/2},\\ \label{sscb}
&&E^2=\mathbf{P}^2 + 2\sqrt{6}\pi^2 T_2 l_{11}^3\mid\mathbf{\Lambda}_1\mid \left(P^2_{\theta}+P^2_{\tilde{\theta}}\right)^{1/2}.\ea

Finally, let us give the semiclassical limit of the membrane solution (\ref{sg1}), which is
\ba\label{sg1SCl} \sigma_{scl}(r)&=&
\left\{\frac{32(4\pi^2T_2 l_{11}^3)^2\left[\mathbf{\Lambda}_1^2 E^2 - \left(\mathbf{\Lambda}_1.\mathbf{P}\right)^2\right]}{27 E^2\left(P^2_{\theta}+P^2_{\tilde{\theta}}\right)}\right\}^{1/4}(l\Delta r)^{1/2}
\\ \nl &\times & F_D^{(3)}\left(1/2;-1/2,-1/2,1/2;3/2;-\frac{\Delta r}{2l},-\frac{\Delta r}{4l},-\frac{\Delta r}{6l}\right)
\\ \nl &=& \left\{\frac{32(4\pi^2T_2 l_{11}^3)^2\left[\mathbf{\Lambda}_1^2 E^2 - \left(\mathbf{\Lambda}_1.\mathbf{P}\right)^2\right]}{27 E^2\left(P^2_{\theta}+P^2_{\tilde{\theta}}\right)}\right\}^{1/4}(l\Delta r)^{1/2}\\ \nl &\times &
\left(1+\frac{\Delta r}{2l}\right)^{1/2}\left(1+\frac{\Delta r}{4l}\right)^{1/2} 
\left(1+\frac{\Delta r}{6l}\right)^{-1/2}
\\ \nl &&F_D^{(3)}\left(1;-1/2,-1/2,1/2,;3/2;\frac{1}{1+\frac{2l}{\Delta r}}, \frac{1}{1+\frac{4l}{\Delta r}}, \frac{1}{1+\frac{6l}{\Delta r}}\right).\ea

\subsection{Second type of membrane embedding}
Let us consider membrane, which is extended along the radial direction $r$ and rotates in the planes defined by the angles $\theta$ and $\tilde{\theta}$, with angular momenta $P_{\theta}$ and $P_{\tilde{\theta}}$. Now we want to have nontrivial wrapping along $X^6$ and $X^9$. The embedding parameters in $X^6$ and $X^9$ have to be chosen in such a way that the constraints (\ref{01}), (\ref{02}) and the equalities $\mathcal{P}^2_\mu=constants$ are identically satisfied.
It turns out that the angular momenta $P_{\theta}$ and $P_{\tilde{\theta}}$ must be equal, and the constants of the motion $\mathcal{P}^2_\mu$ are identically zero for this case. In addition, we want the membrane to move along $X^0$ and $X^I$ with constant energy $E$ and constant momenta $P_{I}$ respectively. All this leads to the following ansatz:
\ba\nl
&&X^0\equiv t=\Lambda_0^0\tau,\h X^I=\Lambda_0^I\tau ,\h X^4\equiv r(\s),
\\ \label{A2} &&X^6\equiv \theta=\Lambda_0^6\tau+\Lambda_1^6\delta + \Lambda_2^6\sigma, \h X^9\equiv\tilde{\theta}=\Lambda_0^6\tau-(\Lambda_1^6\delta + \Lambda_2^6\sigma).\ea

The background felt by the membrane is the same as in (\ref{b1}), but the metric induced on the membrane worldvolume is different and is given by
\ba\nl &&G_{00}=-l_{11}^{2}\left[(\Lambda_0^0)^2-\mathbf{\Lambda}_0^2-(\Lambda_0^+)^2 B^2\right],
\h G_{11}=4l_{11}^{2}(\Lambda_1^6)^2 A^2,
\\ \nl &&G_{12}=4l_{11}^{2}\Lambda_1^6\Lambda_2^6 A^2,\h G_{22}=l_{11}^{2}\left[\frac{r'^2}{C^2} + 4(\Lambda_2^6)^2 A^2\right].\ea

For the present case, the Lagrangian (\ref{old}) reduces to
\ba\nl &&\mathcal{L}^{A}(\sigma) =\frac{1}{4\lambda^0}\left(K_{rr}r'^2 - V\right),\h
K_{rr}=-(4\lambda^0 T_2 l_{11}^2)^2(\Lambda_1^6)^2\frac{A^2}{C^2},\\ \nl
&&V=U=l_{11}^{2}\left[(\Lambda_0^0)^2-\mathbf{\Lambda}_0^2 -(\Lambda_0^+)^2 B^2\right].\ea
The turning points of the effective one-dimensional periodic motion, obtained
from the remaining constraint (\ref{00e})
\ba\nl K_{rr}r'^2 + V=0,\ea
are given by
\ba\nl &&r_{min}=3l,\h  
r_{max}=r_1=l\left(2\sqrt{1+\frac{3v_0^2}{l^2(\Lambda_0^+)^2}}+1\right)>3l,
\\ \label{r'II} &&r_2=-l\left(2\sqrt{1+\frac{3v_0^2}{l^2(\Lambda_0^+)^2}}-1\right)<0, \h v_0^2=(\Lambda_0^0)^2-\mathbf{\Lambda}_0^2.\ea

Replacing the above expressions for $K_{rr}$ and $V$ in (\ref{gf}), we obtain the membrane solution:
\ba\nl &&\sigma(r)=\int_{3l}^{r}\left[-\frac{K_{rr}(t)}{V(t)}\right]^{1/2}dt= 
\frac{32\lambda^0 T_2 l_{11}\Lambda_1^6}{\Lambda_0^+}
\left[\frac{l^3\Delta r}{\left(r_1-3l\right)\left(3l-r_2\right)}\right]^{1/2}\times
\\ \label{s2} &&F_D^{(4)}\left(1/2;-1,-1/2,1/2,1/2;3/2;-\frac{\Delta r}{2l},-\frac{\Delta r}{4l},
-\frac{\Delta r}{3l-r_2},\frac{\Delta r}{r_1-3l}\right).\ea
The normalization condition (\ref{nc}) leads to the following relation between the parameters
\ba\nl &&\frac{16\lambda^0 T_2 l_{11}\Lambda_1^6 l^{3/2}}{\Lambda_0^+\left(3l-r_2\right)^{1/2}}
F_D^{(3)}\left(1/2;-1,-1/2,1/2;1;-\frac{\Delta r_1}{2l},-\frac{\Delta r_1}{4l},-\frac{\Delta r_1}{3l-r_2}\right)
\\ \nl &&=\frac{16\lambda^0 T_2 l_{11}\Lambda_1^6 l^{3/2}}{\Lambda_0^+\left(3l-r_2\right)^{1/2}}
\left(1+\frac{\Delta r_1}{2l}\right)\left(1+\frac{\Delta r_1}{4l}\right)^{1/2} \left(1+\frac{\Delta r_1}{3l-r_2}\right)^{-1/2}\\ \label{nc2}
&&\times F_D^{(3)}\left(1/2;-1,-1/2,1/2,;1;\frac{1}{1+\frac{2l}{\Delta r_1}}, \frac{1}{1+\frac{4l}{\Delta r_1}}, 
\frac{1}{1+\frac{3l-r_2}{\Delta r_1}}\right)=1.\ea
In the case under consideration, the conserved quantities are $E$, $\mathbf{P}$ and $P_{\theta}=P_{\tilde{\theta}}$.
By using (\ref{cmom}), we derive the following result for $P_{\theta}=P_{\tilde{\theta}}$
\ba\nl &&P_{\theta}=P_{\tilde{\theta}}=
\frac{8\pi^2T_2 l_{11}^3\Lambda_1^6 l^{5/2}}{3\left(3l-r_2\right)^{1/2}}\Delta r_1
F_D^{(3)}\left(3/2;-1,-3/2,1/2;2;-\frac{\Delta r_1}{2l},-\frac{\Delta r_1}{4l},-\frac{\Delta r_1}{3l-r_2}\right)
\\ \nl &&=\frac{8\pi^2T_2 l_{11}^3\Lambda_1^6 l^{5/2}}{3\left(3l-r_2\right)^{1/2}}\Delta r_1
\left(1+\frac{\Delta r_1}{2l}\right)\left(1+\frac{\Delta r_1}{4l}\right)^{3/2} \left(1+\frac{\Delta r_1}{3l-r_2}\right)^{-1/2}\\ \label{cmom2}
&&\times F_D^{(3)}\left(1/2;-1,-3/2,1/2,;2;\frac{1}{1+\frac{2l}{\Delta r_1}}, \frac{1}{1+\frac{4l}{\Delta r_1}}, 
\frac{1}{1+\frac{3l-r_2}{\Delta r_1}}\right).\ea

In the semiclassical limit, (\ref{nc2}) and (\ref{cmom2}) reduce to
\ba\nl (\Lambda_0^+)^2=\frac{8\sqrt{3}}{\pi}\lambda^0 T_2 l_{11}\Lambda_1^6\left[(\Lambda_0^0)^2-\mathbf{\Lambda}_0^2\right]^{1/2},
\h P_{\theta}=P_{\tilde{\theta}}=\frac{16\pi T_2 l_{11}^3\Lambda_1^6 }{\sqrt{3}(\Lambda_0^+)^3}
\left[(\Lambda_0^0)^2-\mathbf{\Lambda}_0^2\right]^{3/2}.\ea
From here and (\ref{EP1}), one obtains the relation
\ba\label{EPc2} E^2=\mathbf{P}^2+3^{5/3}(2\pi T_2 l_{11}^3\Lambda_1^6)^{2/3}P_{\theta}^{4/3}.\ea
In the particular case when $\mathbf{P}=0$, (\ref{EPc2}) coincides with the energy-charge relation
$E\sim K^{2/3}$, first obtained for $G_2$-manifolds in \cite{27}.
For the given embedding (\ref{A2}), the semiclassical limit of the membrane solution (\ref{s2}) is as follows
\ba\nl &&\sigma_{scl}(r)=8\pi^{1/3}\left(\frac{2\pi^2 T_2 l_{11}^3\Lambda_1^6}{9P_{\theta}}\right)^{2/3}
\left(l^3\Delta r\right)^{1/2}
F_D^{(2)}\left(1/2;-1,-1/2;3/2;-\frac{\Delta r}{2l},-\frac{\Delta r}{4l}\right)
\\ \label{SCls2} &&=8\pi^{1/3}\left(\frac{2\pi^2 T_2 l_{11}^3\Lambda_1^6}{9P_{\theta}}\right)^{2/3}
\left(l^3\Delta r\right)^{1/2} \left(1+\frac{\Delta r}{2l}\right)\left(1+\frac{\Delta r}{4l}\right)^{1/2}
\\ \nl &&\times F_D^{(2)}\left(1;-1,-1/2;3/2;\frac{1}{1+\frac{2l}{\Delta r}}, 
\frac{1}{1+\frac{4l}{\Delta r}}\right).\ea

\subsection{Third type of membrane embedding}
Again, we want the membrane to move in the flat, four dimensional part of the eleven dimensional background metric (\ref{11db}), with constant energy $E$ and constant momenta $P_{I}$. On the curved part of the metric, the membrane is extended along the radial coordinate $r$, rotates in the plane given by the angle $\psi_+= \psi+\tilde{\psi}$, and is wrapped along the angular coordinate $\psi_-=\psi-\tilde{\psi}$. This membrane configuration is given by
\ba\nl
&&X^0\equiv t=\Lambda_0^0\tau,\h X^I=\Lambda_0^I\tau,\h X^4\equiv r(\s),
\\ \label{A3} &&\psi_+=\Lambda_0^+\tau,\h \psi_-=\Lambda_1^-\delta + \Lambda_2^-\sigma,
\h \psi_\pm = \psi\pm\tilde{\psi}.\ea
In this case, the target space metric seen by the membrane is
\ba\nl &&g_{00}\equiv g_{tt}=-l_{11}^{2},\h g_{IJ}=l_{11}^{2}\delta_{IJ},
\h g_{44}\equiv g_{rr}=\frac{l_{11}^{2}}{C^2(r)},
\\ \label{b2} &&g_{++}=l_{11}^{2}\left(\frac{2l}{3}\right)^2 C^2(r),\h g_{--}=l_{11}^{2}D^2(r).\ea
Hence, in the notations introduced in (\ref{sLA}), we have $\mu=(0,I,+,-)$, $a=4\equiv r$.
Now, the metric induced on the membrane worldvolume is 
\ba\nl &&G_{00}=-l_{11}^{2}\left[(\Lambda_0^0)^2-\mathbf{\Lambda}_0^2 
-(\Lambda_0^+)^2 \left(\frac{2l}{3}\right)^2 C^2\right],\\ \nl 
&&G_{11}=l_{11}^{2}(\Lambda_1^-)^2 D^2,\h G_{12}=l_{11}^{2}\Lambda_1^-\Lambda_2^- D^2,\h G_{22}=l_{11}^{2}\left[(\Lambda_2^-)^2 D^2 + \frac{r'^2}{C^2}\right].\ea
The constraints (\ref{01}), (\ref{02}) are satisfied identically, and $\mathcal{P}^2_\mu\equiv 0$. 
The Lagrangian (\ref{old}) takes the form
\ba\nl &&\mathcal{L}^{A}(\sigma) =\frac{1}{4\lambda^0}\left(K_{rr}r'^2 - V\right),\h
K_{rr}=-(2\lambda^0 T_2 l_{11}^2\Lambda_1^-)^2\frac{D^2}{C^2},\\ \nl
&&V=U= l_{11}^{2}\left[(\Lambda_0^0)^2-\mathbf{\Lambda}_0^2 
-(\Lambda_0^+)^2 \left(\frac{2l}{3}\right)^2 C^2\right].\ea
The turning points, obtained from (\ref{00e}), read
\ba\nl &&r_{min}=3l,\h  
r_{max}=r_1=l\sqrt{1+\frac{8}{1-\frac{9v_0^2}{4l^2(\Lambda_0^+)^2}}}>3l,
\\ \nl &&r_2=-l\sqrt{1+\frac{8}{1-\frac{9v_0^2}{4l^2(\Lambda_0^+)^2}}}<0, \h v_0^2=(\Lambda_0^0)^2-\mathbf{\Lambda}_0^2.\ea

For the present embedding, we derive the following membrane solution
\ba\label{s3} &&\sigma(r)=\int_{3l}^{r}\left[-\frac{K_{rr}(t)}{V(t)}\right]^{1/2}dt= 
\frac{2\lambda^0 T_2 l_{11}\Lambda_1^-}{\left[(\Lambda_0^+)^2 \left(\frac{2l}{3}\right)^2-v_0^2\right]^{1/2}}\left[\frac{2^7 l^5\Delta r}
{3\left(r_1-3l\right) \left(3l-r_2\right)}\right]^{1/2}\times
\\ \nl &&F_D^{(6)}\left(1/2;-1,-1,-1,1/2,1/2,1/2;3/2;-\frac{\Delta r}{2l},-\frac{\Delta r}{3l},-\frac{\Delta r}{4l},-\frac{\Delta r}{6l},-\frac{\Delta r}{3l-r_2},\frac{\Delta r}{r_1-3l}\right).\ea
The normalization condition (\ref{nc}) leads to
\ba\label{nc3} &&\frac{\lambda^0 T_2 l_{11}\Lambda_1^-}{\left[(\Lambda_0^+)^2 \left(\frac{2l}{3}\right)^2-v_0^2\right]^{1/2}}\left[\frac{2^7 l^5}
{3\left(3l-r_2\right)}\right]^{1/2}\times
\\ \nl &&F_D^{(5)}\left(1/2;-1,-1,-1,1/2,1/2;1;-\frac{\Delta r_1}{2l},-\frac{\Delta r_1}{3l},-\frac{\Delta r_1}{4l},-\frac{\Delta r_1}{6l},-\frac{\Delta r_1}{3l-r_2}\right)=
\\ \nl &&\frac{\lambda^0 T_2 l_{11}\Lambda_1^-}{\left[(\Lambda_0^+)^2 \left(\frac{2l}{3}\right)^2-v_0^2\right]^{1/2}}\left[\frac{2^7 l^5}
{3\left(3l-r_2\right)}\right]^{1/2}\times
\\ \nl &&\left(1+\frac{\Delta r_1}{2l}\right)\left(1+\frac{\Delta r_1}{3l}\right)
\left(1+\frac{\Delta r_1}{4l}\right)\left(1+\frac{\Delta r_1}{6l}\right)^{-1/2}
\left(1+\frac{\Delta r_1}{3l-r_2}\right)^{-1/2}\times 
\\ \nl &&F_D^{(5)}\left(1/2;-1,-1,-1,1/2,1/2,;1;\frac{1}{1+\frac{2l}{\Delta r_1}}, \frac{1}{1+\frac{3l}{\Delta r_1}}, \frac{1}{1+\frac{4l}{\Delta r_1}}, \frac{1}{1+\frac{6l}{\Delta r_1}},\frac{1}{1+\frac{3l-r_2}{\Delta r_1}}\right)=1.\ea
The computation of the conserved momentum $P_+ \equiv P_{\psi_+}$ in accordance with (\ref{cmom}) gives
\ba\label{cmom3} &&P_+ = \frac{\pi^2 T_2 l_{11}^3\Lambda_0^+\Lambda_1^-}{\left[(\Lambda_0^+)^2 \left(\frac{2l}{3}\right)^2-v_0^2\right]^{1/2}}\left[\frac{2^5 l^7}
{3^3\left(3l-r_2\right)}\right]^{1/2}\times
\\ \nl &&\Delta r_1 F_D^{(3)}\left(3/2;-1,-1/2,1/2;2;-\frac{\Delta r_1}{3l},-\frac{\Delta r_1}{6l},-\frac{\Delta r_1}{3l-r_2}\right)=
\\ \nl &&\frac{\pi^2 T_2 l_{11}^3\Lambda_0^+\Lambda_1^-}{\left[(\Lambda_0^+)^2 \left(\frac{2l}{3}\right)^2-v_0^2\right]^{1/2}}\left[\frac{2^5 l^7}
{3^3\left(3l-r_2\right)}\right]^{1/2}\times
\\ \nl &&\Delta r_1\left(1+\frac{\Delta r_1}{3l}\right)\left(1+\frac{\Delta r_1}{6l}\right)^{1/2}
\left(1+\frac{\Delta r_1}{3l-r_2}\right)^{-1/2}\times 
\\ \nl &&F_D^{(3)}\left(1/2;-1,-1/2,1/2;2;\frac{1}{1+\frac{3l}{\Delta r_1}},
\frac{1}{1+\frac{6l}{\Delta r_1}},\frac{1}{1+\frac{3l-r_2}{\Delta r_1}}\right).\ea
Let us note that for the embedding (\ref{A3}), the momentum $P_{\psi_-}$ is zero.

Going to the semiclassical limit $r_1\to\infty$, which in the case under consideration leads to
$9v_0^2/[4l^2(\Lambda_0^+)^2]\to 1_-$, one obtains that (\ref{nc3}) and (\ref{cmom3}) reduce to
\ba\nl \Lambda_0^+\left[1-\frac{9v_0^2}{4l^2(\Lambda_0^+)^2}\right]^{3/2}=2\lambda^0 T_2 l_{11}\Lambda_1^-l,
\h P_+ = \frac{2^{5/2}\pi^2T_2 l_{11}^3\Lambda_1^-l^3}{9\left[1-\frac{9v_0^2}{4l^2(\Lambda_0^+)^2}\right]^{3/2}}.\ea
These two equalities, together with (\ref{EP1}), give the following relation between the energy and the conserved momenta
\ba\label{EPc3} E^2= \mathbf{P}^2 + \frac{9}{2l^2}P_+^2 - (6\pi^2T_2 l_{11}^3\Lambda_1^-)^{2/3}P_+^{4/3}.\ea
In the particular case when $\mathbf{P}=0$, (\ref{EPc3}) can be rewritten as
\ba\nl E=\frac{3}{\sqrt{2}l}P_+\sqrt{1-\left(\frac{4\sqrt{2}\pi^2T_2 l_{11}^3\Lambda_1^-l^3}{9P_+}\right)^{2/3}}.\ea
Expanding the square root and neglecting the higher order terms, one derives energy-charge relation of the type
$E-K\sim K^{1/3}$, first found for backgrounds of $G_2$-holonomy in \cite{27}.

Now, let us write down the semiclassical limit of our membrane solution (\ref{s3}):
\ba\label{s3scl} &&\sigma_{scl}(r)=
\frac{\pi^2T_2l_{11}^3\Lambda_1^-}{P_+}\left(\frac{2^7 l^5}{3^3}\right)^{1/2}\times \\ \nl 
&&\Delta r^{1/2} F_D^{(4)}\left(1/2;-1,-1,-1,1/2;3/2;-\frac{\Delta r}{2l},-\frac{\Delta r}{3l},-\frac{\Delta r}{4l},-\frac{\Delta r}{6l}\right)=
\\ \nl &&\frac{\pi^2T_2l_{11}^3\Lambda_1^-}{P_+}\left(\frac{2^7 l^5}{3^3}\right)^{1/2}
\Delta r^{1/2}\left(1+\frac{\Delta r}{2l}\right)\left(1+\frac{\Delta r}{3l}\right)
\left(1+\frac{\Delta r}{4l}\right)\left(1+\frac{\Delta r}{6l}\right)^{-1/2}\times
\\ \nl &&F_D^{(4)}\left(1;-1,-1,-1,1/2;3/2;\frac{1}{1+\frac{2l}{\Delta r}}, \frac{1}{1+\frac{3l}{\Delta r}}, \frac{1}{1+\frac{4l}{\Delta r}}, \frac{1}{1+\frac{6l}{\Delta r}}\right).\ea

\subsection{Forth type of membrane embedding}
Let us consider membrane configuration given by the following ansatz:
\ba\nl
&&X^0\equiv t=\Lambda_0^0\tau+\frac{1}{\Lambda_0^0}\left[\left(\mathbf{\Lambda}_0.\mathbf{\Lambda}_1\right)\delta + \left(\mathbf{\Lambda}_0.\mathbf{\Lambda}_2\right)\sigma\right],
\h X^I=\Lambda_0^I\tau + \Lambda_1^I\delta + \Lambda_2^I\sigma, 
\\ \label{A4} &&X^4\equiv r(\s),\h \psi_+=\Lambda_0^+\tau,\h \psi_-=\Lambda_0^-\tau ,
\h \psi_{\pm}=\psi\pm\tilde{\psi}.\ea
It is analogous to (\ref{A1}), but now the rotations are in the planes defined by the angles $\psi_{\pm}=\psi\pm\tilde{\psi}$ instead of $\theta$ and $\tilde{\theta}$.

The background felt by the membrane is as given in (\ref{b2}).
However, the metric induced on the membrane worldvolume is different and it is the following
\ba\nl &&G_{00}=-l_{11}^{2}\left[(\Lambda_0^0)^2-\mathbf{\Lambda}_0^2 
-(\Lambda_0^+)^2 \left(\frac{2l}{3}\right)^2 C^2 - (\Lambda_0^-)^2 D^2 \right],\\ \nl 
&&G_{11}=l_{11}^{2}M_{11},\h G_{12}=l_{11}^{2}M_{12},\h G_{22}=l_{11}^{2}\left[M_{22} + \frac{r'^2}{C^2}\right],\ea
where $M_{ij}$ are defined in (\ref{DM}). The constraints (\ref{01}), (\ref{02}) are identically satisfied, and the constants of the motion $\mathcal{P}^2_\mu$ are given by (\ref{cm1}).
The Lagrangian (\ref{old}) now takes the form
\ba\nl &&\mathcal{L}^{A}(\sigma) =\frac{1}{4\lambda^0}\left(K_{rr}r'^2 - V\right),\h
K_{rr}=-(2\lambda^0 T_2 l_{11}^2)^2\frac{M_{11}}{C^2},\\ \nl
&&V=(2\lambda^0 T_2 l_{11}^2)^2 \det M_{ij} + 
l_{11}^{2}\left[(\Lambda_0^0)^2-\mathbf{\Lambda}_0^2 
-(\Lambda_0^+)^2 \left(\frac{2l}{3}\right)^2 C^2 - (\Lambda_0^-)^2 D^2 \right].\ea

Let us first consider the particular case when $\Lambda_0^-=0$, i.e. $\psi=\tilde{\psi}$. 
The turning points obtained from the constraint (\ref{00e}) now are
\ba\nl r_{min}=3l,\h  
r_{max}=r_1=l\sqrt{1+\frac{8}{1-\frac{9u_0^2}{4l^2(\Lambda_0^+)^2}}}>3l,
\h r_2=-l\sqrt{1+\frac{8}{1-\frac{9u_0^2}{4l^2(\Lambda_0^+)^2}}}<0,\ea
where $u_0^2$ is introduced in (\ref{u02}). By using (\ref{gf}), one arrives at the following membrane solution
\ba\label{s4} &&\sigma(r)=\int_{3l}^{r}\left[-\frac{K_{rr}(t)}{U(t)}\right]^{1/2}dt= 
\frac{2\lambda^0 T_2 l_{11}}{\left[(\Lambda_0^+)^2 \left(\frac{2l}{3}\right)^2-u_0^2\right]^{1/2}}\left[\frac{2^7 l^3M_{11}\Delta r}{3\left(r_1-3l\right) \left(3l-r_2\right)}\right]^{1/2}
\\ \nl &&\times F_D^{(5)}\left(1/2;-1,-1,1/2,1/2,1/2;3/2;-\frac{\Delta r}{2l},-\frac{\Delta r}{4l},-\frac{\Delta r}{6l},-\frac{\Delta r}{3l-r_2},\frac{\Delta r}{r_1-3l}\right).\ea
The normalization condition (\ref{nc}) now gives
\ba\label{nc4} &&\frac{\lambda^0 T_2 l_{11}}{\left[(\Lambda_0^+)^2 \left(\frac{2l}{3}\right)^2-u_0^2\right]^{1/2}}\left[\frac{2^7 l^3M_{11}}{3\left(3l-r_2\right)}\right]^{1/2}\times
\\ \nl &&F_D^{(4)}\left(1/2;-1,-1,1/2,1/2;1;-\frac{\Delta r_1}{2l},-\frac{\Delta r_1}{4l},-\frac{\Delta r_1}{6l},-\frac{\Delta r_1}{3l-r_2}\right)=
\\ \nl &&\frac{\lambda^0 T_2 l_{11}}{\left[(\Lambda_0^+)^2 \left(\frac{2l}{3}\right)^2-u_0^2\right]^{1/2}}\left[\frac{2^7 l^3M_{11}}{3\left(3l-r_2\right)}\right]^{1/2}\times
\\ \nl &&\left(1+\frac{\Delta r_1}{2l}\right)\left(1+\frac{\Delta r_1}{4l}\right)\left(1+\frac{\Delta r_1}{6l}\right)^{-1/2}\left(1+\frac{\Delta r_1}{3l-r_2}\right)^{-1/2}\times 
\\ \nl &&F_D^{(4)}\left(1/2;-1,-1,1/2,1/2,;1;\frac{1}{1+\frac{2l}{\Delta r_1}}, \frac{1}{1+\frac{4l}{\Delta r_1}}, \frac{1}{1+\frac{6l}{\Delta r_1}},\frac{1}{1+\frac{3l-r_2}{\Delta r_1}}\right)=1.\ea
In accordance with (\ref{cmom}), we derive for the conserved momentum $P_+\equiv P_{\psi_+}$ the expression ($P_-\equiv P_{\psi_-}=0$ as a consequence of $\Lambda_0^-=0$):
\ba\nl &&P_+ =\frac{\pi^2 T_2 l_{11}^3\Lambda_0^+}{\left[(\Lambda_0^+)^2 \left(\frac{2l}{3}\right)^2-u_0^2\right]^{1/2}}\left[\frac{2^5 l^5 M_{11}}{3^3\left(3l-r_2\right)}\right]^{1/2}\times
\\ \nl &&\Delta r_1 F_D^{(2)}\left(3/2;-1/2,1/2;2;-\frac{\Delta r_1}{6l},-\frac{\Delta r_1}{3l-r_2}\right)=\\ \nl
&&\frac{\pi^2 T_2 l_{11}^3\Lambda_0^+}{\left[(\Lambda_0^+)^2 \left(\frac{2l}{3}\right)^2-u_0^2\right]^{1/2}}\left[\frac{2^5 l^5 M_{11}}{3^3\left(3l-r_2\right)}\right]^{1/2}
\Delta r_1 \left(1+\frac{\Delta r_1}{6l}\right)^{1/2}\left(1+\frac{\Delta r_1}{3l-r_2}\right)^{-1/2}\times 
\\ \label{cmom4} &&F_D^{(2)}\left(1/2;-1/2,1/2;2;\frac{1}{1+\frac{6l}{\Delta r_1}},\frac{1}{1+\frac{3l-r_2}{\Delta r_1}}\right).\ea

In the semiclassical limit, (\ref{nc4}) and (\ref{cmom4}) simplify to 
\ba\nl \pi\Lambda_0^+\left[1-\frac{9u_0^2}{4l^2(\Lambda_0^+)^2}\right]=2^{3/2}3\lambda^0 T_2 l_{11}M_{11}^{1/2},
\h P_+ = \frac{2^{7/2}\pi T_2 l_{11}^3l^2M_{11}^{1/2}}{3\left[1-\frac{9u_0^2}{4l^2(\Lambda_0^+)^2}\right]}.\ea
Taking also into account (\ref{EP1}), we obtain the following {\it fourth} order algebraic equation for $E^2$ as a function of $\mathbf{P}$ and $P_+$
\ba\nl &&\left\{E^2\left[E^2-\mathbf{P}^2-(3/l)^2P_+^2\right] - (2\pi^2 T_2 l_{11}^3)^2\left\{\left(\mathbf{\Lambda}_1\times\mathbf{\Lambda}_2\right)^2 E^2 - \left[\left(\mathbf{\Lambda}_1\times\mathbf{\Lambda}_2\right)\times\mathbf{P}\right]^2\right\}\right\}^2 
\\ \label{Eg4} &&-2^7(3\pi T_2 l_{11}^3)^2E^2\left[\mathbf{\Lambda}_1^2 E^2 - \left(\mathbf{\Lambda}_1.\mathbf{P}\right)^2\right]P_{+}^2 = 0.\ea
Let us consider a few simple cases.
When $\Lambda_0^I=0$ and $\Lambda_2^I=c\Lambda_1^I$, (\ref{Eg4}) reduces to
\ba\label{p41} E^2=(3/l)^2P_+^2 + 2^{7/2}3\pi T_2 l_{11}^3\mid\mathbf{\Lambda}_1\mid P_{+},\ea
or
\ba\nl
E=\frac{3}{l}P_+\sqrt{1+ \frac{2^{7/2}\pi T_2 l_{11}^3l^2\mid\mathbf{\Lambda}_1\mid}{3P_+}}.\ea
Expanding the square root and neglecting the higher order terms, one derives energy-charge relation of the type
$E-K\sim const$.
If we impose only the conditions $\Lambda_0^I=0$, (\ref{Eg4}) gives
\ba\label{p42} E^2 =(2\pi^2 T_2 l_{11}^3)^2\left(\mathbf{\Lambda}_1\times\mathbf{\Lambda}_2\right)^2  + 
(3/l)^2P_+^2+2^{7/2}3\pi T_2 l_{11}^3\mid\mathbf{\Lambda}_1\mid P_{+}.\ea
If we take $\Lambda_0^I\ne 0$, $\Lambda_2^I=c\Lambda_1^I$, (\ref{Eg4}) simplifies to
\ba\nl E^2\left\{\left[E^2-\mathbf{P}^2-(3/l)^2P_+^2\right]^2 - 2^7(3\pi T_2 l_{11}^3)^2 \mathbf{\Lambda}_1^2P_{+}^2\right\}
+2^7(3\pi T_2 l_{11}^3)^2\left(\mathbf{\Lambda}_1.\mathbf{P}\right)^2P_{+}^2=0 ,\ea
which is {\it third} order algebraic equation for $E^2$. Suppose that $\mathbf{\Lambda}_1$ and $\mathbf{P}$ are orthogonal to each other, i.e. $\left(\mathbf{\Lambda}_1.\mathbf{P}\right)=0$. Then, the above relation becomes
\ba\label{p43} E^2=\mathbf{P}^2 + (3/l)^2P_+^2+2^{7/2}3\pi T_2 l_{11}^3\mid\mathbf{\Lambda}_1\mid P_{+}.\ea
Finally, we give the semiclassical limit of the membrane solution (\ref{s4})
\ba\nl &&\sigma_{scl}(r)=2\pi^2T_2 l_{11}^3\left(\frac{4l}{3}\right)^{3/2}\left[\mathbf{\Lambda}_1^2  - \frac{1}{E^2}\left(\mathbf{\Lambda}_1.\mathbf{P}\right)^2\right]^{1/2}\frac{\Delta r^{1/2}}{P_+}
\\ \nl &&\times F_D^{(3)}\left(1/2;-1,-1,1/2;3/2;-\frac{\Delta r}{2l},-\frac{\Delta r}{4l},
-\frac{\Delta r}{6l}\right)
\\ \nl &&=2\pi^2T_2 l_{11}^3\left(\frac{4l}{3}\right)^{3/2}\left[\mathbf{\Lambda}_1^2  - \frac{1}{E^2}\left(\mathbf{\Lambda}_1.\mathbf{P}\right)^2\right]^{1/2}\frac{\Delta r^{1/2}}{P_+}
\left(1+\frac{\Delta r}{2l}\right)\left(1+\frac{\Delta r}{4l}\right)
\left(1+\frac{\Delta r}{6l}\right)^{-1/2}\\ \nl &&\times
F_D^{(3)}\left(1;-1,-1,1/2;3/2;\frac{1}{1+\frac{2l}{\Delta r}}, \frac{1}{1+\frac{4l}{\Delta r}}, \frac{1}{1+\frac{6l}{\Delta r}}\right).\ea

Now, we turn to the case $\Lambda_0^-\ne 0$, when the solutions of the equation $r'=0$ are
\ba\nl &&r_{min}=3l,\h  
r_{max}=r_1=\frac{l}{\sqrt{2}}\sqrt{1+u^2-\Lambda^2}\sqrt{1+ \sqrt{1-\frac{4(u^2-9\Lambda^2)}{(1+u^2-\Lambda^2)^2}}},
\\ \nl &&r_2=\frac{l}{\sqrt{2}}\sqrt{1+u^2-\Lambda^2}\sqrt{1- \sqrt{1-\frac{4(u^2-9\Lambda^2)}{(1+u^2-\Lambda^2)^2}}},
\\ \nl &&r_3=-\frac{l}{\sqrt{2}}\sqrt{1+u^2-\Lambda^2}\sqrt{1+ \sqrt{1-\frac{4(u^2-9\Lambda^2)}{(1+u^2-\Lambda^2)^2}}},
\\ \nl &&r_4=-\frac{l}{\sqrt{2}}\sqrt{1+u^2-\Lambda^2}\sqrt{1- \sqrt{1-\frac{4(u^2-9\Lambda^2)}{(1+u^2-\Lambda^2)^2}}},
\\ \nl &&u^2=\left(\frac{3u_0}{l\Lambda_0^-}\right)^2,\h \Lambda^2=\left(2\frac{\Lambda_0^+}{\Lambda_0^-}\right)^2.\ea
Correspondingly, we obtain the following solution for $\s(r)$:
\ba\nl &&\sigma(r)=\int_{3l}^{r}\left[-\frac{K_{rr}(t)}{U(t)}\right]^{1/2}dt= 
\frac{\lambda^0 T_2 l_{11}}{\Lambda_0^-}\left[\frac{2^9 3 l^3M_{11}\Delta r}
{\left(r_1-3l\right) \left(3l-r_2\right) \left(3l-r_3\right) \left(3l-r_4\right)}\right]^{1/2}\times
\\ \label{s4g} &&F_D^{(7)}\left(1/2;-1,-1,1/2,1/2,1/2,1/2,1/2;3/2;\right.
\\ \nl &&-\left.\frac{\Delta r}{2l},-\frac{\Delta r}{4l},-\frac{\Delta r}{6l},-\frac{\Delta r}{3l-r_2},-\frac{\Delta r}{3l-r_3},-\frac{\Delta r}{3l-r_4},
\frac{\Delta r}{r_1-3l}\right).\ea
For the normalization condition, we derive the result
\ba\label{nc4g} &&\frac{\lambda^0 T_2 l_{11}}{\Lambda_0^-}\left[\frac{2^7 3 l^3M_{11}}
{\left(3l-r_2\right) \left(3l-r_3\right) \left(3l-r_4\right)}\right]^{1/2}\times
\\ \nl &&F_D^{(6)}\left(1/2;-1,-1,1/2,1/2,1/2,1/2;1;\right.
\\ \nl &&-\left.\frac{\Delta r_1}{2l},-\frac{\Delta r_1}{4l},-\frac{\Delta r_1}{6l},-\frac{\Delta r_1}{3l-r_2},-\frac{\Delta r_1}{3l-r_3},-\frac{\Delta r_1}{3l-r_4}\right)=
\\ \nl &&\frac{\lambda^0 T_2 l_{11}}{\Lambda_0^-}\left[\frac{2^7 3 l^3M_{11}}
{\left(3l-r_2\right) \left(3l-r_3\right) \left(3l-r_4\right)}\right]^{1/2}
\left(1+ \frac{\Delta r}{2l}\right)\left(1+ \frac{\Delta r}{4l}\right)\left(1+ \frac{\Delta r}{6l}\right)^{-1/2}\times
\\ \nl &&\left(1+ \frac{\Delta r}{3l-r_2}\right)^{-1/2} \left(1+ \frac{\Delta r}{3l-r_3}\right)^{-1/2} \left(1+ \frac{\Delta r}{3l-r_4}\right)^{-1/2}\times
\\ \nl &&F_D^{(6)}\left(1/2;-1,-1,1/2,1/2,1/2,1/2;1;\right.
\\ \nl &&\left.\frac{1}{1+\frac{2l}{\Delta r_1}}, \frac{1}{1+\frac{4l}{\Delta r_1}}, \frac{1}{1+\frac{6l}{\Delta r_1}},\frac{1}{1+\frac{3l-r_2}{\Delta r_1}} ,\frac{1}{1+\frac{3l-r_3}{\Delta r_1}} ,\frac{1}{1+\frac{3l-r_4}{\Delta r_1}}\right)=1.\ea
The computation of the conserved quantities $P_+$ and $P_-$ gives
\ba\label{cmom4g+} &&P_+=\pi^2T_2 l_{11}^3\frac{\Lambda_0^+}{\Lambda_0^-}\left[\frac{2^5 l^5 M_{11}}
{3\left(3l-r_2\right) \left(3l-r_3\right) \left(3l-r_4\right)}\right]^{1/2}\Delta r_1\times
\\ \nl &&F_D^{(4)}\left(3/2;-1/2,1/2,1/2,1/2;2;-\frac{\Delta r_1}{6l},-\frac{\Delta r_1}{3l-r_2},-\frac{\Delta r_1}{3l-r_3},-\frac{\Delta r_1}{3l-r_4}\right)=
\\ \nl &&\pi^2T_2 l_{11}^3\frac{\Lambda_0^+}{\Lambda_0^-}\left[\frac{2^5 l^5 M_{11}}
{3\left(3l-r_2\right) \left(3l-r_3\right) \left(3l-r_4\right)}\right]^{1/2}\times
\\ \nl &&\Delta r_1\left(1+ \frac{\Delta r_1}{6l}\right)^{1/2}\left(1+ \frac{\Delta r_1}{3l-r_2}\right)^{-1/2} \left(1+ \frac{\Delta r_1}{3l-r_3}\right)^{-1/2} \left(1+ \frac{\Delta r_1}{3l-r_4}\right)^{-1/2}\times
\\ \nl &&F_D^{(4)}\left(1/2;-1/2,1/2,1/2,1/2;2; \frac{1}{1+\frac{6l}{\Delta r_1}},\frac{1}{1+\frac{3l-r_2}{\Delta r_1}} ,\frac{1}{1+\frac{3l-r_3}{\Delta r_1}} ,\frac{1}{1+\frac{3l-r_4}{\Delta r_1}}\right),\ea

\ba\label{cmom4g-} &&P_-=\pi^2T_2 l_{11}^3\left[\frac{2^7 3 l^7 M_{11}}
{\left(3l-r_2\right) \left(3l-r_3\right) \left(3l-r_4\right)}\right]^{1/2}\times
\\ \nl &&F_D^{(7)}\left(1/2;-1,-2,-1,1/2,1/2,1/2,1/2;1;\right.
\\ \nl &&-\left.\frac{\Delta r_1}{2l},-\frac{\Delta r_1}{3l},-\frac{\Delta r_1}{4l},-\frac{\Delta r_1}{6l},-\frac{\Delta r_1}{3l-r_2},-\frac{\Delta r_1}{3l-r_3},-\frac{\Delta r_1}{3l-r_4}\right)=
\\ \nl &&\pi^2T_2 l_{11}^3\left[\frac{2^7 3 l^7 M_{11}}
{\left(3l-r_2\right) \left(3l-r_3\right) \left(3l-r_4\right)}\right]^{1/2}\times
\\ \nl &&\left(1+ \frac{\Delta r_1}{2l}\right)\left(1+ \frac{\Delta r_1}{3l}\right)^2\left(1+ \frac{\Delta r_1}{4l}\right)\left(1+ \frac{\Delta r_1}{6l}\right)^{-1/2}\times
\\ \nl &&\left(1+ \frac{\Delta r_1}{3l-r_2}\right)^{-1/2} \left(1+ \frac{\Delta r_1}{3l-r_3}\right)^{-1/2} \left(1+ \frac{\Delta r_1}{3l-r_4}\right)^{-1/2}\times
\\ \nl &&F_D^{(7)}\left(1/2;-1,-2,-1,1/2,1/2,1/2,1/2;1;\right. 
\\ \nl &&\left.\frac{1}{1+\frac{2l}{\Delta r_1}},\frac{1}{1+\frac{3l}{\Delta r_1}},\frac{1}{1+\frac{4l}{\Delta r_1}},\frac{1}{1+\frac{6l}{\Delta r_1}},\frac{1}{1+\frac{3l-r_2}{\Delta r_1}} ,\frac{1}{1+\frac{3l-r_3}{\Delta r_1}},\frac{1}{1+\frac{3l-r_4}{\Delta r_1}}\right).\ea

Let us now take the semiclassical limit $r_1\to\infty$.  In this limit, (\ref{nc4g}), (\ref{cmom4g+}) and (\ref{cmom4g-}) reduce correspondingly to
\ba\nl &&\Lambda_0^- = 3\lambda^0T_2l_{11}M_{11}^{1/2},\h
P_+ = \frac{4}{3}\pi^2T_2l_{11}^3l^2M_{11}^{1/2}\frac{\Lambda_0^+}{\Lambda_0^-},\\ \nl
&&P_- = \frac{1}{6}\pi^2T_2l_{11}^3l^2M_{11}^{1/2} \left[\left(\frac{3u_0}{l\Lambda_0^-}\right)^2- \left(2\frac{\Lambda_0^+}{\Lambda_0^-}\right)^2\right].\ea
These equalities, together with (\ref{EP1}), lead to the following relation between the energy $E$ and the conserved charges $\mathbf{P}$, $P_+$ and $P_-$:
\ba\nl &&\left\{E^2\left[E^2-\mathbf{P}^2-(3/2l)^2 P_+^2\right] - (2\pi^2 T_2 l_{11}^3)^2\left\{\left(\mathbf{\Lambda}_1\times\mathbf{\Lambda}_2\right)^2 E^2 - \left[\left(\mathbf{\Lambda}_1\times\mathbf{\Lambda}_2\right)\times\mathbf{P}\right]^2\right\}\right\}^2 
\\ \label{Egg4} &&-(6\pi^2 T_2 l_{11}^3)^2E^2\left[\mathbf{\Lambda}_1^2 E^2 - \left(\mathbf{\Lambda}_1.\mathbf{P}\right)^2\right]P_{-}^2 = 0.\ea
We remind the reader that the  above relation is only valid for $P_-\ne 0$, whereas we can always set $\mathbf{P}$ or $P_+$ equal to zero. Below, we give a few simple solutions of (\ref{Egg4}).

Choosing $\Lambda_0^I=0$ and $\Lambda_2^I=c\Lambda_1^I$, one obtains
\ba\label{p41g} E^2=(3/2l)^2P_+^2 + 6\pi^2 T_2 l_{11}^3\mid\mathbf{\Lambda}_1\mid P_{-},\ea
which can be rewritten as
\ba\nl
E=\frac{3}{2l}P_+\sqrt{1+ \frac{8\pi^2 T_2 l_{11}^3l^2\mid\mathbf{\Lambda}_1\mid P_{-}}{3P_+^2}}.\ea
Expanding the square root and neglecting the higher order terms, one arrives at
\ba\nl E= \frac{3}{2l}P_+ + 2\pi^2 T_2 l_{11}^3l\mid\mathbf{\Lambda}_1\mid \frac{P_{-}}{P_{+}}.\ea
If only the conditions $\Lambda_0^I=0$ are imposed, (\ref{Egg4}) gives
\ba\label{p42g} E^2 =(2\pi^2 T_2 l_{11}^3)^2\left(\mathbf{\Lambda}_1\times\mathbf{\Lambda}_2\right)^2  + 
(3/2l)^2P_+^2 + 6\pi^2 T_2 l_{11}^3\mid\mathbf{\Lambda}_1\mid P_{-}.\ea
If we choose $\Lambda_0^I\ne 0$, $\Lambda_2^I=c\Lambda_1^I$, then (\ref{Egg4}) simplifies to a {\it third} order algebraic equation for $E^2$
\ba\nl E^2\left\{\left[E^2-\mathbf{P}^2-(3/2l)^2P_+^2\right]^2 - (6\pi^2 T_2 l_{11}^3)^2\mathbf{\Lambda}_1^2P_{-}^2\right\}
+(6\pi^2 T_2 l_{11}^3)^2\left(\mathbf{\Lambda}_1.\mathbf{P}\right)^2P_{-}^2=0 .\ea
If $\left(\mathbf{\Lambda}_1.\mathbf{P}\right)=0$, the above relation reduces to
\ba\label{p43g} E^2=\mathbf{P}^2 + (3/2l)^2P_+^2+6\pi^2 T_2 l_{11}^3\mid\mathbf{\Lambda}_1\mid P_{-}.\ea

Finally, let us write down the semiclassical limit of the membrane solution (\ref{s4g}):
\ba\nl &&\sigma_{scl}(r)= \left(\frac{2^8\pi^2T_2 l_{11}^3l}{3^4P_-}\right)^{1/2}\left[\mathbf{\Lambda}_1^2  - \frac{1}{E^2}\left(\mathbf{\Lambda}_1.\mathbf{P}\right)^2\right]^{1/4}\Delta r^{1/2}
\\ \nl &&\times F_D^{(4)}\left(1/2;-1,1,-1,1/2;3/2;-\frac{\Delta r}{2l},-\frac{\Delta r}{3l}, 
-\frac{\Delta r}{4l},-\frac{\Delta r}{6l}\right)
\\ \nl &&=\left(\frac{2^8\pi^2T_2 l_{11}^3l}{3^4P_-}\right)^{1/2}\left[\mathbf{\Lambda}_1^2  - \frac{1}{E^2}\left(\mathbf{\Lambda}_1.\mathbf{P}\right)^2\right]^{1/4}
\\ \nl &&\times \Delta r^{1/2}
\left(1+\frac{\Delta r}{2l}\right)\left(1+\frac{\Delta r}{3l}\right)^{-1}\left(1+\frac{\Delta r}{4l}\right)
\left(1+\frac{\Delta r}{6l}\right)^{-1/2}\\ \nl &&\times
F_D^{(4)}\left(1;-1,1,-1,1/2;3/2;\frac{1}{1+\frac{2l}{\Delta r}}, \frac{1}{1+\frac{3l}{\Delta r}}, \frac{1}{1+\frac{4l}{\Delta r}}, \frac{1}{1+\frac{6l}{\Delta r}}\right).\ea

Concluding this section, we note that more membrane solutions are given in Appendix B. The reason is that although   different, they exhibit the same semiclassical behavior as some of the solutions described here. Namely, they lead to the same dependence of the energy on the conserved charges in this limit.

\setcounter{equation}{0}
\section{Concluding remarks}
In this paper, we considered the membrane dynamics on a manifold with exactly known metric of $G_2$-holonomy in M-theory. More precisely, we obtained exact {\it rotating} membrane solutions and explicit expressions for the energy $E$ and the other momenta (charges), which are conserved due to the presence of background isometries. They were given in terms of the hypergeometric functions of many variables $F_D^{(n)}(a;b_1,\ldots,b_n;c;z_1,\ldots,z_n)$, where for the different membrane configurations considered, $n$ varies from two to seven.

In connection with the dual four dimensional $\mathcal{N}=1$ gauge theory, we investigated the semiclassical limit of the conserved quantities and received different types of relations between them. In particular, we reproduced the energy-charge relations $E\sim K^{1/2}$, $E\sim K^{2/3}$ and $E-K\sim K^{1/3}$, first found for rotating membranes on backgrounds of $G_2$-holonomy in this limit in \cite{27}. Moreover, we found examples of more complicated dependence of the energy on the charges. The most general cases considered, lead to algebraic equations of third or even forth order for the $E^2$ as a function of up to five conserved momenta. Presumably, these may correspond to operators of more general type in the dual field theory. Also, they could be connected with the lack of conformal invariance.

As already observed in \cite{27} for rotating membranes on $G_2$ manifolds, one may have the same energy-charge relations in the limits of {\it small} and {\it large} charges. Such are $E\sim K^{1/2}$ and $E\sim K^{2/3}$ \cite{27}. Let us give an example, which confirms this observation. For large charges, according to (\ref{p1}), the following equality holds:
\ba\nl E_l=2(\sqrt{3}\pi^2 T_2 l_{11}^3\mid\mathbf{\Lambda}_1\mid)^{1/2} (P_{\theta}^{l})^{1/2}.\ea  
On the other hand, taking the small charge limit in the expression (\ref{cmom1}) for $P_{\theta}$, which corresponds to $\Delta r_1\to 0$, one obtains the relation
\ba\nl E_s=2(2\pi^2 T_2 l_{11}^3\mid\mathbf{\Lambda}_1\mid)^{1/2} (P_{\theta}^{s})^{1/2}.\ea 
Hence, in both cases, we have the same $E\sim K^{1/2}$ behavior. As a consequence, the ratio of the two energies is given by: 
\ba\nl E_l/E_s=(3/4)^{1/4}(P_{\theta}^{l}/P_{\theta}^{s})^{1/2}.\ea

Here, we did not investigate the limit of {\it small} conserved charges. However, the {\it exact} expressions for all quantities which we are interested in, are written in two forms: one appropriate for considering the large charges limit, and the other - for small ones. That is why, the last limit can be always done.

For comparison, we now give the known results about the different energy-charge relations in the semiclassical limit, for membranes moving on other curved M-theory backgrounds. So far, such relations have been obtained for the following target spaces: $AdS_p\times S^q$, $AdS_4\times Q^{1,1,1}$, warped $AdS_5\times M^6$, and 11-dimensional $AdS$-black hole \cite{6}, \cite{10}, \cite{27}, \cite{BRR04}-\cite{0508}. If we denote the conserved angular momentum on the $AdS$-part of the metric with $S$ and on the other part with $J$, the known expressions for $E(S,J)$ are as follows. 
\paragraph{1. On the $AdS_p\times S^q$ backgrounds} \cite{6}, \cite{10}, \cite{27}, \cite{BRR04}-\cite{0508}
\ba\nl &&E-S\sim S^{1/3},\h E-S=c_1 S^{1/3}+c_2\frac{J^2}{S^{1/3}}+\ldots,\h E-S\sim \ln\frac{S}{c}, 
\\ \nl &&E=J+\ldots,\h E-c_1J= \frac{c_2}{J^3}\sum_{a,b=1}^{d}c_{ab}J_aJ_b + \ldots,\h E=c_1S+c_2J^2. \ea

\paragraph{2. On the $AdS_4\times Q^{1,1,1}$ background} \cite{27}
\ba\nl E-S\sim \ln\frac{S}{c},\h E=J+\ldots.\ea

\paragraph{3. On the warped $AdS_5\times M^6$ background} \cite{27}
\ba\nl E-S\sim \ln\frac{S}{c},\h E-J=c+\ldots.\ea

\paragraph{4. On the 11-dimensional $AdS$-black hole background} \cite{27}
\ba\nl E-cS\sim S^3.\ea

It seems to us that an interesting task is to find rotating string configurations in type IIA theory in ten dimensions, which reproduce the energy-charge relations obtained here, for rotating membranes on an
eleven dimensional background with $G_2$ holonomy. This problem is under investigation \cite{PLB}, and now we give an example of such string solution. 

As explained in section 2, the reduction to ten dimensions of the M-theory background (\ref{11db}) is given by (\ref{10db}), which describes a D6-brane wrapping the ${\bf S}^3$ in the deformed conifold geometry. Let us consider the following string embedding in (\ref{10db}):
\ba\nl X^0=\Lambda_0^0\tau,\h X^I=\Lambda_0^I\tau ,\h r= r(\s),
\h\theta_1=\Lambda_0^{\theta_1}\tau, \h \theta_2=\Lambda_0^{\theta_2}\tau,
\h \psi_1=\phi_1=\phi_2=0.\ea
This ansatz corresponds to string, which is extended along the radial direction $r$, rotates in the planes defined by the angles $\theta_1$ and $\theta_2$ with angular momenta $P_{\theta_1}$ and $P_{\theta_2}$, and moves along $X^0$ and $X^I$ with constant energy $E$ and constant momenta $P_{I}$ respectively. It can be shown that for large conserved charges, the dependence of the energy $E$ on $P_{I}$, $P_{\theta_1}$ and $P_{\theta_2}$ is 
\ba\nl E^2=\mathbf{P}^2 + const\left(P^2_{\theta_1}+P^2_{\theta_2}\right)^{1/2}.\ea
Thus, this string configuration has the same semiclassical behavior as the membrane in (\ref{sscb}). 

To our knowledge, none of the energy-charge relations obtained here for membranes moving on a $G_2$ manifold correspond to usual relations, coming from operators in the dual $\mathcal{N}=1$ gauge theory. The most plausible explanation is that the Kaluza-Klein modes are not fully decoupled from the pure SYM theory excitations. In this respect, a good idea for exploration of the problem is the one proposed in \cite{GN05}. In this article, the SL(3,R) deformations of a type IIB background based on D5-branes that is conjectured to be dual to $\mathcal{N}=1$ SYM \cite{MN00} are studied.  It is argued that this deformation only affects the Kaluza-Klein sector of the dual field theory and helps decoupling the Kaluza-Klein dynamics from the pure gauge dynamics. Recently, 
evidences for the above prediction have been given in \cite{BDR11216}. In this paper, semiclassical strings on the deformed Maldacena-Nunez background \cite{GN05} are studied and the results are compared with those obtained previously for the undeformed case \cite{ponstal}.  It was observed there that the string energies increase due to the deformation, which is interpreted as a proof for better decoupling of the Kaluza-Klein modes in the deformed theory. This is in accordance with \cite{GN05}, where it was conjectured that the sectors in which the deformation is decoupled, should correspond to pure gauge theory effects. As an additional evidence for the above idea, the authors of \cite{BDR11216} consider a particular string configuration, for which the string energy is independent of the deformation. The articles \cite{GN05} and \cite{BDR11216} give us the line for further investigations in this direction. First, by performing $TsT$ transformation \cite{F05}, one obtains the deformed eleven dimensional background. Second, find rotating membrane solutions in this new background. Third, compare the energies of the membranes moving on the original and on the deformed backgrounds and so on. The same could be done for strings in type IIA theory in ten dimensions, which reproduce the energy-charge relations obtained for rotating membranes. Then, a natural question is whether the dimensional reduction and the deformation commute? We hope to be able to report our results on these problems soon.

\vspace*{.5cm} 

{\bf Acknowledgments} \vspace*{.2cm}

The author would like to thank R.C. Rashkov for the valuable discussion. 

This work is supported by NSFB grant under contract $\Phi1412/04$.

\newpage
\appendix
\setcounter{equation}{0}
\section{Hypergeometric functions $F_{D}^{(n)}$}

Here, we give some properties of the hypergeometric functions of many variables 
$F_{D}^{(n)}$ used in our calculations. By definition \cite{PBM-III}, for $|z_j|<1$,
\ba\nl F_D^{(n)}(a;b_1,\ldots,b_n;c;z_1,\ldots,z_n)=
\sum_{k_1,\ldots,k_n=0}^{\infty}\frac{(a)_{k_1+\ldots+k_n}(b_1)_{k_1}\ldots(b_n)_{k_n}}
{(c)_{k_1+\ldots+k_n}}\frac{z_1^{k_1}\ldots z_n^{k_n}}{k_1!\ldots k_n!},\ea
where
\ba\nl (a)_k = \frac{\Gamma(a+k)}{\Gamma(a)},\ea
and $\Gamma(z)$ is the Euler's $\Gamma$-function.
In particular, $F_D^{(1)}(a;b;c;z)= \mbox{}_2F_{1}(a,b;c;z)$ is the Gauss' hypergeometric function,
and $F_D^{(2)}(a;b_1,b_2;c;z_1,z_2)= F_{1}(a,b_1,b_2;c;z_1,z_2)$ is one of the hypergeometric 
functions of two variables.
\ba\nl
{\bf 1.} &&F_D^{(n)}(a;b_1,\ldots,b_i,\ldots,b_j,\ldots,b_n;c;z_1,\ldots,z_i,\ldots,z_j,\ldots,z_n)= 
\\ \nl &&F_D^{(n)}(a;b_1,\ldots,b_j,\ldots,b_i,\ldots,b_n;c;z_1,\ldots,z_j,\ldots,z_i,\ldots,z_n).
\\ \nl 
{\bf 2.} &&F_D^{(n)}(a;b_1,\ldots,b_n;c;z_1,\ldots,z_n)= 
\\ \nl &&\prod_{i=1}^{n}\left(1-z_i\right)^{-b_i}
F_D^{(n)}\left(c-a;b_1,\ldots,b_n;c;\frac{z_1}{z_1-1},\ldots,\frac{z_n}{z_n-1}\right).
\\ \nl
{\bf 3.} &&F_D^{(n)}(a;b_1,\ldots,b_{i-1},b_i,b_{i+1},\ldots,b_n;c;z_1,\ldots,z_{i-1},1,z_{i+1},\ldots,z_n)=
\\ \nl &&\frac{\Gamma(c)\Gamma(c-a-b_i)}{\Gamma(c-a)\Gamma(c-b_i)}
F_D^{(n-1)}(a;b_1,\ldots,b_{i-1},b_{i+1},\ldots,b_n;c-b_i;z_1,\ldots,z_{i-1},z_{i+1},\ldots,z_n).
\\ \nl
{\bf 4.} &&F_D^{(n)}(a;b_1,\ldots,b_{i-1},b_i,b_{i+1},\ldots,b_n;c;z_1,\ldots,z_{i-1},0,z_{i+1},\ldots,z_n)=
\\ \nl &&F_D^{(n-1)}(a;b_1,\ldots,b_{i-1},b_{i+1},\ldots,b_n;c;z_1,\ldots,z_{i-1},z_{i+1},\ldots,z_n).
\\ \nl
{\bf 5.} &&F_D^{(n)}(a;b_1,\ldots,b_{i-1},0,b_{i+1},\ldots,b_n;c;z_1,\ldots,z_{i-1},z_i,z_{i+1},\ldots,z_n)=
\\ \nl &&F_D^{(n-1)}(a;b_1,\ldots,b_{i-1},b_{i+1},\ldots,b_n;c;z_1,\ldots,z_{i-1},z_{i+1},\ldots,z_n).
\\ \nl
{\bf 6.} &&F_D^{(n)}(a;b_1,\ldots,b_i,\ldots,b_j,\ldots,b_n;c;z_1,\ldots,z_i,\ldots,z_i,\ldots,z_n)=
\\ \nl &&F_D^{(n-1)}(a;b_1,\ldots,b_i+b_j,\ldots,b_n;c;z_1,\ldots,z_i,\ldots,z_n).
\\ \nl
{\bf 7.} &&F_D^{(2n+1)}(a;a-c+1,b_2,b_2,\ldots,b_{2n},b_{2n};c;-1,z_2,-z_2\ldots,z_{2n},-z_{2n})=
\\ \nl &&\frac{\Gamma(a/2)\Gamma(c)}{2\Gamma(a)\Gamma(c-a/2)}
F_D^{(n)}(a/2;b_2,\ldots,b_{2n};c-a/2;z_2^2,\ldots,z_{2n}^2).
\\ \nl
{\bf 8.} &&F_D^{(2n+1)}\left(c-a;a-c+1,b_2,b_2,\ldots,b_{2n},b_{2n};c;\right. \\ \nl  &&\left.1/2,-\frac{z_2}{1-z_2},\frac{z_2}{1+z_2},\ldots,-\frac{z_{2n}}{1-z_{2n}},\frac{z_{2n}}{1+z_{2n}}\right)=
\\ \nl &&\frac{\Gamma(a/2)\Gamma(c)}{2^{c-a}\Gamma(a)\Gamma(c-a/2)}
F_D^{(n)}\left(c-a;b_2,\ldots,b_{2n};c-a/2;-\frac{z_2^2}{1-z_2^2},\ldots,-\frac{z_{2n}^2}{1-z_{2n}^2}\right).
\\ \nl
{\bf 9.} &&F_D^{(2)}\left(a;b,b;c;z,-z\right)= 
\mbox{}_3F_{2}\left(\matrix{a/2,(a+1)/2,b \\ \nl c/2,(c+1)/2;z^2}\right).\cr\ea

\newpage
\setcounter{equation}{0}
\section{More solutions}
Here, we give other exact rotating membrane solutions and explicit expressions 
for the corresponding conserved quantities, which lead to the same dependence 
of the energy on the charges in the semiclassical limit, as part of 
those described in section 4.

\subsection{Fifth type of membrane embedding}
Now, consider membrane, which moves with constant energy $E$ and momenta $P_I$ and is extended along the radial direction $r$. Also, it rotates in the plane defined by the angle $\phi_+=\phi+\tilde{\phi}$. In addition, the membrane is wrapped along the angular coordinates $\psi=\tilde{\psi}$ and $\phi_-=\phi-\tilde{\phi}$. This configuration corresponds to the following ansatz, for which the constraints (\ref{01}), (\ref{02}) are identically satisfied, and $\mathcal{P}^2_\mu\equiv 0$
\footnote{This is also true for all other embeddings further considered.}:
\ba\nl &&X^0\equiv t = \Lambda_0^0\tau, \h X^I=\Lambda_0^I\tau ,\h X^4\equiv r(\s),\\ \nl
&&\psi=\tilde{\psi} = \Lambda_1^{\psi}\left(\delta + \frac{\Lambda_2^-}{\Lambda_1^-}\sigma\right),
\h \phi_- = \Lambda_1^-\delta + \Lambda_2^-\sigma,\h \phi_+ = \Lambda_0^+\tau ;
\h \phi_{\pm}=\phi\pm\tilde{\phi}.\ea
The background felt by the membrane in this case, as well as in all other cases considered below, is
\ba\nl &&g_{00}\equiv g_{tt}=-l_{11}^{2},\h g_{IJ}=l_{11}^{2}\delta_{IJ},
\h g_{44}\equiv g_{rr}=\frac{l_{11}^{2}}{C^2(r)},
\\ \nl &&g_{\psi\psi}=l_{11}^{2}\left(\frac{4l}{3}\right)^2 C^2(r),
\h g_{--}=l_{11}^{2}A^2(r),\h g_{++}=l_{11}^{2}B^2(r).\ea
Therefore, in the notations introduced in (\ref{sLA}), we have 
$\mu=(0,I,\psi,-,+)$, $a=4\equiv r$.
The Lagrangian (\ref{old}) takes the form
\ba\nl &&\mathcal{L}^{A}(\sigma) =\frac{1}{4\lambda^0}\left(K_{rr}r'^2 - V\right),\h
K_{rr}=-(2\lambda^0 T_2 l_{11}^2)^2\left[(\Lambda_1^-)^2\frac{A^2}{C^2}+ \left(\frac{4l}{3}\right)^2(\Lambda_1^{\psi})^2\right],\\ \nl
&&V=U= l_{11}^{2}\left[(\Lambda_0^0)^2-\mathbf{\Lambda}_0^2 
-(\Lambda_0^+)^2 B^2\right].\ea
The turning points defined by $r'=0$ coincide with those given in (\ref{r'II}).
The solution (\ref{gf}) now reads
\ba\nl &&\sigma(r)=\int_{3l}^{r}\left[-\frac{K_{rr}(t)}{U(t)}\right]^{1/2}dt= 
4\lambda^0T_2l_{11}\frac{\Lambda_1^-}{\Lambda_0^+}
\left[\frac{\prod_{\alpha=1}^{3}\left(3l-w_{\alpha}\right)}{\left(r_1-3l\right) \left(3l-r_2\right)}\right]^{1/2}
\Delta r^{1/2}\times\\ \nl 
&&F_D^{(5)}\left(1/2;1/2,1/2,-1/2,-1/2,-1/2;3/2;\right.\\ \nl &&\left.
-\frac{\Delta r}{3l-r_2},\frac{\Delta r}{r_1-3l},-\frac{\Delta r}{3l-w_1},-\frac{\Delta r}{3l-w_2},-\frac{\Delta r}{3l-w_3}\right),\ea
where $w_{\alpha}(\Lambda_1^{\psi})$ ($\alpha=1,2,3$) are the zeros of the polynomial 
\ba\nl t^3-lt^2-l^2\left[1-\left(\frac{8\Lambda_1^{\psi}}{\sqrt{3}}\right)^2\right] t
+l^3\left[1-3\left(\frac{8\Lambda_1^{\psi}}{\sqrt{3}}\right)^2\right]=
(t-w_1)(t-w_2)(t-w_3).\ea
The normalization condition (\ref{nc}) leads to
\ba\nl &&2\lambda^0T_2l_{11}\frac{\Lambda_1^-}{\Lambda_0^+}
\left[\frac{\prod_{\alpha=1}^{3}\left(3l-w_{\alpha}\right)}{3l-r_2}\right]^{1/2}\times\\ \nl 
&&F_D^{(4)}\left(1/2;1/2,-1/2,-1/2,-1/2;1;
-\frac{\Delta r_1}{3l-r_2},-\frac{\Delta r_1}{3l-w_1},-\frac{\Delta r_1}{3l-w_2},-\frac{\Delta r_1}{3l-w_3}\right)=
\\ \nl &&2\lambda^0T_2l_{11}\frac{\Lambda_1^-}{\Lambda_0^+}
\left[\frac{\prod_{\alpha=1}^{3}\left(3l-w_{\alpha}\right)}{3l-r_2}\right]^{1/2}
\left(1+\frac{\Delta r_1}{3l-r_2}\right)^{-1/2} \prod_{\alpha=1}^{3}\left(1+\frac{\Delta r_1}{3l-w_{\alpha}}\right)^{1/2}\times \\ \nl
&&F_D^{(4)}\left(1/2;1/2,-1/2,-1/2,-1/2;1; \frac{1}{1+\frac{3l-r_2}{\Delta r_1}} ,\frac{1}{1+\frac{3l-w_1}{\Delta r_1}} ,\frac{1}{1+\frac{3l-w_2}{\Delta r_1}},\frac{1}{1+\frac{3l-w_3}{\Delta r_1}}\right)=1.\ea
In accordance with (\ref{cmom}), we derive the following expression  for the conserved momentum $P_+\equiv P_{\phi_+}$:
\ba\nl &&P_+=\frac{l}{3}\pi^{2}T_2l_{11}^{3}\Lambda_1^-
\left[\frac{\prod_{\alpha=1}^{3}\left(3l-w_{\alpha}\right)}{3l-r_2}\right]^{1/2}\Delta r_1\times\\ \nl 
&&F_D^{(5)}\left(3/2;-1,1/2,-1/2,-1/2,-1/2;2;-\frac{\Delta r_1}{4l},
-\frac{\Delta r_1}{3l-r_2},-\frac{\Delta r_1}{3l-w_1},-\frac{\Delta r_1}{3l-w_2},-\frac{\Delta r_1}{3l-w_3}\right)
\\ \nl &&=\frac{l}{3}\pi^{2}T_2l_{11}^{3}\Lambda_1^- 
\left[\frac{\prod_{\alpha=1}^{3}\left(3l-w_{\alpha}\right)}{3l-r_2}\right]^{1/2}
\\ \nl &&\times\Delta r_1 \left(1+\frac{\Delta r_1}{4l}\right)\left(1+\frac{\Delta r_1}{3l-r_2}\right)^{-1/2} \prod_{\alpha=1}^{3}\left(1+\frac{\Delta r_1}{3l-w_{\alpha}}\right)^{1/2} \\ \nl
&&\times F_D^{(5)}\left(1/2;-1,1/2,-1/2,-1/2,-1/2;2;\right.
\\ \nl &&\left.\frac{1}{1+\frac{4l}{\Delta r_1}}, \frac{1}{1+\frac{3l-r_2}{\Delta r_1}} ,\frac{1}{1+\frac{3l-w_1}{\Delta r_1}} ,\frac{1}{1+\frac{3l-w_2}{\Delta r_1}},\frac{1}{1+\frac{3l-w_3}{\Delta r_1}}\right).\ea

Taking the semiclassical limit\footnote{In this limit $w_{\alpha}$ remain finite.}, we obtain the following dependence of the energy on $\mathbf{P}$ and $P_+$:
\ba\nl E^2=\mathbf{P}^2 + 3^{5/3}(\pi T_2l_{11}^3\Lambda_1^-)^{2/3}P_+^{4/3},\ea
which is of the same type as (\ref{EPc2}). The semiclassical limit of the solution $\s(r)$ is given by:
\ba\nl &&\s_{scl}(r)=2\pi^{1/3}\left(\frac{\pi^2 T_2 l_{11}^3\Lambda_1^-}{9P_+}\right)^{2/3}
\left[\prod_{\alpha=1}^{3}\left(3l-w_{\alpha}\right)\right]^{1/2}\Delta r^{1/2}\times
\\ \nl &&F_D^{(3)}\left(1/2;-1/2,-1/2,-1/2;3/2;
-\frac{\Delta r}{3l-w_1},-\frac{\Delta r}{3l-w_2},-\frac{\Delta r}{3l-w_3}\right)
\\ \nl &&= 2\pi^{1/3}\left(\frac{\pi^2 T_2 l_{11}^3\Lambda_1^-}{9P_+}\right)^{2/3}
\left[\prod_{\alpha=1}^{3}\left(3l-w_{\alpha}\right)\right]^{1/2}\Delta r^{1/2} 
\prod_{\alpha=1}^{3}\left(1+\frac{\Delta r}{3l-w_{\alpha}}\right)^{1/2}
\\ \nl &&\times F_D^{(3)}\left(1;-1/2,-1/2,-1/2;3/2; \frac{1}{1+\frac{3l-w_1}{\Delta r}},
\frac{1}{1+\frac{3l-w_2}{\Delta r}},\frac{1}{1+\frac{3l-w_3}{\Delta r}}\right).\ea

\subsection{Sixth type of membrane embedding}
Let us take the following membrane configuration: 
\ba\nl &&X^0\equiv t = \Lambda_0^0\tau, \h X^I=\Lambda_0^I\tau ,\h X^4\equiv r(\s),\\ \nl
&&\psi=\tilde{\psi} = \Lambda_1^{\psi}\left(\delta + \frac{\Lambda_2^+}{\Lambda_1^+}\sigma\right),
\h \phi_+ = \Lambda_1^+\delta + \Lambda_2^+\sigma,\h \phi_- = \Lambda_0^-\tau .\ea
It is similar to the case just considered, but the roles of the angles $\phi_+$ and $\phi_-$ are interchanged.
Although the exact classical expressions for the quantities we are interested in are different 
from those obtained for the previously considered embedding, one arrives at the same semiclassical behavior:
\ba\nl E^2=\mathbf{P}^2 + 3^{5/3}(\pi T_2l_{11}^3\Lambda_1^+)^{2/3}P_-^{4/3}.\ea

\subsection{Seventh type of membrane embedding}
Now, we consider membrane embedding, which corresponds to rotation in the plane given by the angle $\psi=\tilde{\psi}$, and wrapping along $\phi_+$ and $\phi_-$:
\ba\nl &&X^0\equiv t = \Lambda_0^0\tau, \h X^I=\Lambda_0^I\tau ,\h X^4\equiv r(\s),\\ \nl
&&\psi=\tilde{\psi} = \Lambda_0^{\psi}\tau,
\h \phi_- =\Lambda_1^-\delta + \Lambda_2^-\sigma,
\h \phi_+ = \Lambda_1^+\left(\delta + \frac{\Lambda_2^-}{\Lambda_1^-}\sigma\right).\ea
The effective Lagrangian (\ref{old}) now reads
\ba\nl &&\mathcal{L}^{A}(\sigma) =\frac{1}{4\lambda^0}\left(K_{rr}r'^2 - V\right),\h
K_{rr}=-(2\lambda^0 T_2 l_{11}^2)^2\frac{1}{C^2}\left[(\Lambda_1^-)^2A^2+ (\Lambda_1^{+})^2B^2\right],\\ \nl
&&V=U= l_{11}^{2}\left[(\Lambda_0^0)^2-\mathbf{\Lambda}_0^2 
-\left(\frac{4l}{3}\right)^2(\Lambda_0^{\psi})^2 C^2\right].\ea
For the solutions of the equation $r'=0$ one obtains
\ba\nl &&r_{min}=3l,\h  
r_{max}=r_1=l\sqrt{1+\frac{8}{1-\frac{9v_0^2}{16l^2(\Lambda_0^{\psi})^2}}}>3l,
\\ \nl &&r_2=-l\sqrt{1+\frac{8}{1-\frac{9v_0^2}{16l^2(\Lambda_0^{\psi})^2}}}<0, \h v_0^2=(\Lambda_0^0)^2-\mathbf{\Lambda}_0^2.\ea
For the membrane solution (\ref{gf}), we find the following explicit expression
\ba\nl &&\sigma(r)=\int_{3l}^{r}\left[-\frac{K_{rr}(t)}{U(t)}\right]^{1/2}dt= 
2\lambda^0T_2l_{11}\frac{\left[(\Lambda_1^+)^2+(\Lambda_1^-)^2\right]^{1/2}}
{\Lambda_0^{\psi}\left[1-\frac{9v_0^2}{16l^2(\Lambda_0^{\psi})^2}\right]^{1/2}}
\left[\frac{2l\left(3l-w_+\right)\left(3l-w_-\right)}{\left(r_1-3l\right) \left(3l-r_2\right)}\right]^{1/2}
\\ \nl &&\times \Delta r^{1/2}F_D^{(7)}\left(1/2;-1,-1,1/2,1/2,1/2,-1/2,-1/2;3/2;\right.\\ \nl &&\left.-\frac{\Delta r}{2l},-\frac{\Delta r}{4l},-\frac{\Delta r}{6l},
-\frac{\Delta r}{3l-r_2},\frac{\Delta r}{r_1-3l},-\frac{\Delta r}{3l-w_+},-\frac{\Delta r}{3l-w_-}\right),\ea
where $w_{\pm}$ are given by
\ba\nl w_{\pm} = l\left[\frac{(\Lambda_1^+)^2-(\Lambda_1^-)^2}{(\Lambda_1^+)^2+(\Lambda_1^-)^2}
\pm \sqrt{3+\left(\frac{(\Lambda_1^+)^2-(\Lambda_1^-)^2}{(\Lambda_1^+)^2+(\Lambda_1^-)^2}\right)^2}\right].\ea
The normalization condition (\ref{nc}) gives:
\ba\nl &&\lambda^0T_2l_{11}\frac{\left[(\Lambda_1^+)^2+(\Lambda_1^-)^2\right]^{1/2}}
{\Lambda_0^{\psi}\left[1-\frac{9v_0^2}{16l^2(\Lambda_0^{\psi})^2}\right]^{1/2}}
\left[\frac{2l\left(3l-w_+\right)\left(3l-w_-\right)}{3l-r_2}\right]^{1/2}
\\ \nl &&\times F_D^{(6)}\left(1/2;-1,-1,1/2,1/2,-1/2,-1/2;1;\right.\\ \nl &&\left.-\frac{\Delta r_1}{2l},-\frac{\Delta r_1}{4l},-\frac{\Delta r_1}{6l},
-\frac{\Delta r_1}{3l-r_2},-\frac{\Delta r_1}{3l-w_+},-\frac{\Delta r_1}{3l-w_-}\right)
\\ \nl &&=\lambda^0T_2l_{11}\frac{\left[(\Lambda_1^+)^2+(\Lambda_1^-)^2\right]^{1/2}}
{\Lambda_0^{\psi}\left[1-\frac{9v_0^2}{16l^2(\Lambda_0^{\psi})^2}\right]^{1/2}}
\left[\frac{2l\left(3l-w_+\right)\left(3l-w_-\right)}{3l-r_2}\right]^{1/2}
\\ \nl &&\times \left(1+\frac{\Delta r_1}{2l}\right) \left(1+\frac{\Delta r_1}{4l}\right) \left(1+\frac{\Delta r_1}{6l}\right)^{-1/2} 
\\ \nl &&\times\left(1+\frac{\Delta r_1}{3l-r_2}\right)^{-1/2} 
\left(1+\frac{\Delta r_1}{3l-w_+}\right)^{1/2} \left(1+\frac{\Delta r_1}{3l-w_-}\right)^{1/2}
\\ \nl &&\times F_D^{(6)}\left(1/2;-1,-1,1/2,1/2,-1/2,-1/2;1;\right.\\ \nl &&\left. \frac{1}{1+\frac{2l}{\Delta r_1}},\frac{1}{1+\frac{4l}{\Delta r_1}},\frac{1}{1+\frac{6l}{\Delta r_1}}, \frac{1}{1+\frac{3l-r_2}{\Delta r_1}} ,\frac{1}{1+\frac{3l-w_+}{\Delta r_1}} ,\frac{1}{1+\frac{3l-w_-}{\Delta r_1}}\right)=1.\ea
In the case under consideration, the nontrivial conserved quantities are $E$, $\mathbf{P}$ and $P_{\psi}=P_{\tilde{\psi}}$. By using (\ref{cmom}), we derive the following result for $P_{\psi}$
\ba\nl &&P_{\psi}= \pi^2T_2l_{11}^3\frac{\left[(\Lambda_1^+)^2+(\Lambda_1^-)^2\right]^{1/2}}
{3\left[1-\frac{9v_0^2}{16l^2(\Lambda_0^{\psi})^2}\right]^{1/2}}
\left[\frac{(2l)^3\left(3l-w_+\right)\left(3l-w_-\right)}{3l-r_2}\right]^{1/2}\times
\\ \nl &&\Delta r_1 F_D^{(4)}\left(3/2;-1/2,1/2,-1/2,-1/2;2;-\frac{\Delta r_1}{6l},
-\frac{\Delta r_1}{3l-r_2},-\frac{\Delta r_1}{3l-w_+},-\frac{\Delta r_1}{3l-w_-}\right)
\\ \nl &&= \pi^2T_2l_{11}^3\frac{\left[(\Lambda_1^+)^2+(\Lambda_1^-)^2\right]^{1/2}}
{3\left[1-\frac{9v_0^2}{16l^2(\Lambda_0^{\psi})^2}\right]^{1/2}}
\left[\frac{(2l)^3\left(3l-w_+\right)\left(3l-w_-\right)}{3l-r_2}\right]^{1/2}
\\ \nl &&\times \Delta r_1\left(1+\frac{\Delta r_1}{6l}\right)^{1/2} \left(1+\frac{\Delta r_1}{3l-r_2}\right)^{-1/2} \left(1+\frac{\Delta r_1}{3l-w_+}\right)^{1/2} \left(1+\frac{\Delta r_1}{3l-w_-}\right)^{1/2}
\\ \nl &&\times F_D^{(4)}\left(1/2;-1/2,1/2,-1/2,-1/2;2;\frac{1}{1+\frac{6l}{\Delta r_1}}, \frac{1}{1+\frac{3l-r_2}{\Delta r_1}} ,\frac{1}{1+\frac{3l-w_+}{\Delta r_1}} ,\frac{1}{1+\frac{3l-w_-}{\Delta r_1}}\right).\ea

Based on the above expressions, in the semiclassical limit, we obtain:
\ba\nl E^2= \mathbf{P}^2 + \left(\frac{3}{4l}\right)^2P_{\psi}^2 - 
\frac{3}{4}(\pi^2T_2 l_{11}^3)^{2/3}\left[(\Lambda_1^+)^2+(\Lambda_1^-)^2\right]^{1/3}P_{\psi}^{4/3}.\ea
This is the same type of semiclassical behavior as the one in (\ref{EPc3}). For large conserved charges, the solution $\s(r)$ simplifies to 
\ba\nl &&\sigma_{scl}(r) = \frac{16\pi^2T_2l_{11}^3}{9P_{\psi}}\left[(\Lambda_1^+)^2+(\Lambda_1^-)^2\right]^{1/2}
\left[l^3\left(3l-w_+\right)\left(3l-w_-\right)\right]^{1/2}\times 
\\ \nl &&\Delta r^{1/2} F_D^{(5)}\left(1/2;-1,-1,1/2,-1/2,-1/2;3/2;-\frac{\Delta r}{2l}, -\frac{\Delta r}{4l}, -\frac{\Delta r}{6l},-\frac{\Delta r}{3l-w_+},-\frac{\Delta r}{3l-w_-}\right)
\\ \nl &&=  \frac{16\pi^2T_2l_{11}^3}{9P_{\psi}}\left[(\Lambda_1^+)^2+(\Lambda_1^-)^2\right]^{1/2}
\left[l^3\left(3l-w_+\right)\left(3l-w_-\right)\right]^{1/2}\times 
\\ \nl &&\Delta r^{1/2} \left(1+\frac{\Delta r}{2l}\right)\left(1+\frac{\Delta r}{4l}\right) \left(1+\frac{\Delta r}{6l}\right)^{-1/2} 
\left(1+\frac{\Delta r}{3l-w_+}\right)^{1/2} \left(1+\frac{\Delta r}{3l-w_-}\right)^{1/2}\times 
\\ \nl &&F_D^{(5)}\left(1;-1,-1,1/2,-1/2,-1/2;3/2;\frac{1}{1+\frac{2l}{\Delta r}}, \frac{1}{1+\frac{4l}{\Delta r}},\frac{1}{1+\frac{6l}{\Delta r}}, \frac{1}{1+\frac{3l-w_+}{\Delta r}} ,\frac{1}{1+\frac{3l-w_-}{\Delta r}}\right).\ea

\subsection{Eighth type of membrane embedding}
Here, we investigate the following membrane configuration:
\ba\nl &&X^0\equiv t = \Lambda_0^0\tau, \h X^I=\Lambda_0^I\tau ,\h X^4\equiv r(\s),
\\ \nl &&\psi=\tilde{\psi} = \Lambda_1^{\psi}\delta + \Lambda_2^{\psi}\sigma,
\h \phi_- =\Lambda_0^-\tau, \h \phi_+ = \Lambda_0^+\tau.\ea
It describes membrane, rotating in the planes given by the angles $\phi_\pm$, and wrapped along the coordinate $\psi=\tilde{\psi}$.  
In this case, the reduced Lagrangian (\ref{old}) have the form:
\ba\nl &&\mathcal{L}^{A}(\sigma) =\frac{1}{4\lambda^0}\left(K_{rr}r'^2 - V\right),\h
K_{rr}=-(2\lambda^0 T_2 l_{11}^2)^2\left(\frac{4l}{3}\right)^2(\Lambda_1^{\psi})^2,\\ \nl
&&V=U= l_{11}^{2}\left[(\Lambda_0^0)^2-\mathbf{\Lambda}_0^2 
-(\Lambda_0^-)^2 A^2 - (\Lambda_0^+)^2 B^2\right]=
l_{11}^{2}\left[v_0^2 -(\Lambda_0^-)^2 A^2 - (\Lambda_0^+)^2 B^2\right].\ea
By solving the equation $r'=0$ (see (\ref{00e})), one obtains
\ba\nl r_{\pm}= l\left\{\frac{(\Lambda_0^+)^2-(\Lambda_0^-)^2}{(\Lambda_0^+)^2+(\Lambda_0^-)^2}
\pm \sqrt{3+\left[\frac{(\Lambda_0^+)^2-(\Lambda_0^-)^2}{(\Lambda_0^+)^2+(\Lambda_0^-)^2}\right]^2
+ \frac{12v_0^2}{l^2\left[(\Lambda_0^+)^2+(\Lambda_0^-)^2\right]}}\right\}.\ea
Depending on the sign of $\left[(\Lambda_0^+)^2-(\Lambda_0^-)^2\right]$, we have the following three cases.  
\paragraph{1. $(\Lambda_0^+)^2-(\Lambda_0^-)^2=0$}
\ba\nl r_{max}=r_1=l\sqrt{3 + \frac{6v_0^2}{l^2(\Lambda_0^-)^2}},\h r_2=-r_1.\ea
\paragraph{2. $(\Lambda_0^+)^2-(\Lambda_0^-)^2>0$}
\ba\nl &&r_1=l\frac{(\Lambda_0^+)^2-(\Lambda_0^-)^2}{(\Lambda_0^+)^2+(\Lambda_0^-)^2}
\left\{\sqrt{1 + 3\left[\frac{(\Lambda_0^+)^2+(\Lambda_0^-)^2} {(\Lambda_0^+)^2-(\Lambda_0^-)^2}\right]^2
\left(1+ \frac{4v_0^2}{l^2\left[(\Lambda_0^+)^2+(\Lambda_0^-)^2\right]}\right)}+1\right\},
\\ \nl &&r_2=-l\frac{(\Lambda_0^+)^2-(\Lambda_0^-)^2}{(\Lambda_0^+)^2+(\Lambda_0^-)^2}
\left\{\sqrt{1 + 3\left[\frac{(\Lambda_0^+)^2+(\Lambda_0^-)^2} {(\Lambda_0^+)^2-(\Lambda_0^-)^2}\right]^2
\left(1+ \frac{4v_0^2}{l^2\left[(\Lambda_0^+)^2+(\Lambda_0^-)^2\right]}\right)}-1\right\}.\ea
\paragraph{3. $(\Lambda_0^+)^2-(\Lambda_0^-)^2<0$}
\ba\nl &&r_1=l\frac{(\Lambda_0^-)^2-(\Lambda_0^+)^2}{(\Lambda_0^+)^2+(\Lambda_0^-)^2}
\left\{\sqrt{1 + 3\left[\frac{(\Lambda_0^+)^2+(\Lambda_0^-)^2} {(\Lambda_0^+)^2-(\Lambda_0^-)^2}\right]^2
\left(1+ \frac{4v_0^2}{l^2\left[(\Lambda_0^+)^2+(\Lambda_0^-)^2\right]}\right)}-1\right\},
\\ \nl &&r_2=-l\frac{(\Lambda_0^-)^2-(\Lambda_0^+)^2}{(\Lambda_0^+)^2+(\Lambda_0^-)^2}
\left\{\sqrt{1 + 3\left[\frac{(\Lambda_0^+)^2+(\Lambda_0^-)^2} {(\Lambda_0^+)^2-(\Lambda_0^-)^2}\right]^2
\left(1+ \frac{4v_0^2}{l^2\left[(\Lambda_0^+)^2+(\Lambda_0^-)^2\right]}\right)}+1\right\}.\ea
In all these cases, the condition $r_{max}=r_1>3l=r_{min}$ leads to $v_0^2>l^2(\Lambda_0^-)^2$, so we can consider them simultaneously. 

For the present embedding, the membrane solution (\ref{gf}) has the form
\ba\nl &&\sigma(r)=\int_{3l}^{r}\left[-\frac{K_{rr}(t)}{U(t)}\right]^{1/2}dt
\\ \nl &&=\frac{16\lambda^0T_2l_{11}l\Lambda_1^{\psi}}
{\left[(\Lambda_0^+)^2+(\Lambda_0^-)^2\right]^{1/2}\left[3\left(r_1-3l\right) \left(3l-r_2\right)\right]^{1/2}}
\Delta r F_D^{(2)}\left(1;1/2,1/2;2;-\frac{\Delta r}{3l-r_2},\frac{\Delta r}{r_1-3l}\right),\ea
and the normalization condition (\ref{nc}) reads
\ba\nl &&\frac{32\lambda^0T_2l_{11}l\Lambda_1^{\psi}}
{\left[(\Lambda_0^+)^2+(\Lambda_0^-)^2\right]^{1/2}}\left[\frac{\Delta r_1}{3(3l-r_2)}\right]^{1/2}
 F_D^{(1)}\left(1;1/2;3/2;-\frac{\Delta r_1}{3l-r_2}\right)
 \\ \nl &&=\frac{32\lambda^0T_2l_{11}l\Lambda_1^{\psi}}
{\left[(\Lambda_0^+)^2+(\Lambda_0^-)^2\right]^{1/2}}\left[\frac{\Delta r_1}{3(3l-r_2)}\right]^{1/2}
\mbox{}_2F_{1}\left(1,1/2;3/2;-\frac{\Delta r_1}{3l-r_2}\right)
\\ \nl &&=  \frac{32\lambda^0T_2l_{11}l\Lambda_1^{\psi}}
{\left[(\Lambda_0^+)^2+(\Lambda_0^-)^2\right]^{1/2}}\left[\frac{\Delta r_1}{3(3l-r_2)}\right]^{1/2}
\left(1+ \frac{\Delta r_1}{3l-r_2}\right)^{-1/2}
\mbox{}_2F_{1}\left(1/2,1/2;3/2;\frac{1}{1+\frac{3l-r_2}{\Delta r_1}}\right)
\\ \nl &&=\frac{32\lambda^0T_2l_{11}l\Lambda_1^{\psi}}{3^{1/2}\left[(\Lambda_0^+)^2+(\Lambda_0^-)^2\right]^{1/2}}
\arcsin\left(1+\frac{3l-r_2}{\Delta r_1}\right)^{-1/2}=\pi.\ea
According to (\ref{cmom}), the computation of the conserved momenta $P_\pm = P_{\phi_{\pm}}$ gives
\ba\nl &&P_+ = \frac{64\pi T_2l_{11}^3l^2\Lambda_1^{\psi}\Lambda_0^+}
{3^{5/2}\left[(\Lambda_0^+)^2+(\Lambda_0^-)^2\right]^{1/2}}\left(\frac{\Delta r_1^3}{3l-r_2}\right)^{1/2}
F_D^{(2)}\left(2;-1,1/2;5/2;-\frac{\Delta r_1}{4l},-\frac{\Delta r_1}{3l-r_2}\right)
\\ \nl &&= \frac{64\pi T_2l_{11}^3l^2\Lambda_1^{\psi}\Lambda_0^+}
{3^{5/2}\left[(\Lambda_0^+)^2+(\Lambda_0^-)^2\right]^{1/2}}\left(\frac{\Delta r_1^3}{3l-r_2}\right)^{1/2}
\left(1+ \frac{\Delta r_1}{4l}\right)\left(1+ \frac{\Delta r_1}{3l-r_2}\right)^{-1/2}\\ \nl &&\times
F_D^{(2)}\left(1/2;-1,1/2;5/2;\frac{1}{1+\frac{4l}{\Delta r_1}}, \frac{1}{1+\frac{3l-r_2}{\Delta r_1}}\right),\ea
\ba\nl &&P_- = \frac{32\pi T_2l_{11}^3l^3\Lambda_1^{\psi}\Lambda_0^-}
{3^{1/2}\left[(\Lambda_0^+)^2+(\Lambda_0^-)^2\right]^{1/2}}\left(\frac{\Delta r_1}{3l-r_2}\right)^{1/2}
\\ \nl &&\times
F_D^{(3)}\left(1;-1,-1,1/2;3/2;-\frac{\Delta r_1}{2l},-\frac{\Delta r_1}{6l},-\frac{\Delta r_1}{3l-r_2}\right)
\\ \nl &&= \frac{32\pi T_2l_{11}^3l^3\Lambda_1^{\psi}\Lambda_0^-}
{3^{1/2}\left[(\Lambda_0^+)^2+(\Lambda_0^-)^2\right]^{1/2}}\left(\frac{\Delta r_1}{3l-r_2}\right)^{1/2}
\left(1+ \frac{\Delta r_1}{2l}\right)\left(1+ \frac{\Delta r_1}{6l}\right)
\left(1+ \frac{\Delta r_1}{3l-r_2}\right)^{-1/2}\\ \nl &&\times
F_D^{(3)}\left(1/2;-1,-1,1/2;3/2;\frac{1}{1+\frac{2l}{\Delta r_1}}, 
\frac{1}{1+\frac{6l}{\Delta r_1}},\frac{1}{1+\frac{3l-r_2}{\Delta r_1}}\right).\ea

In the semiclassical limit $r_1\to\infty$, the above expressions for the normalization condition and 
$P_\pm$ reduce to:
\ba\nl \frac{8\lambda^0T_2l_{11}l\Lambda_1^{\psi}}{3^{1/2}\left[(\Lambda_0^+)^2+(\Lambda_0^-)^2\right]^{1/2}}=1, \h
P_\pm = \frac{2^{5/2}\pi^2 T_2l_{11}^3l\Lambda_1^{\psi}\Lambda_0^{\pm}}
{3^{1/2}\left[(\Lambda_0^+)^2+(\Lambda_0^-)^2\right]^{3/2}}v_0^2.\ea
Combining these equalities with (\ref{EP1}), one obtains the following relation between $E$, $\mathbf{P}$ and $P_\pm$: 
\ba\nl E^2=\mathbf{P}^2 + \left(128/3\right)^{1/2}\pi^2 T_2 l_{11}^3 l\Lambda_1^{\psi} \left(P^2_{+}+P^2_{-}\right)^{1/2}.\ea
This is the same semiclassical behavior as in (\ref{sscb}). 

Finally, let us write down the semiclassical limit of the solution $\s(r)$ for the present embedding. It is the simplest one, we have been able to obtain in this paper, and is given by:
\ba\nl &&\sigma_{scl}(r) = \frac{\left(P^2_{+}+P^2_{-}\right)^{1/4}}{(2^5 3)^{1/4}
(\pi^2 T_2 l_{11}^3 l\Lambda_1^{\psi})^{1/2}}\Delta r F_D^{(2)}\left(1;1/2,1/2;2;-\frac{\Delta r}{\Delta r_1},
\frac{\Delta r}{\Delta r_1}\right)
\\ \nl &&=\frac{\left(P^2_{+}+P^2_{-}\right)^{1/4}}{(2^5 3)^{1/4}
(\pi^2 T_2 l_{11}^3 l\Lambda_1^{\psi})^{1/2}}\Delta r \hspace{.1cm}
\mbox{}_3F_{2}\left(\matrix{1/2,1,1/2 \\ \nl 1,3/2;\frac{\Delta r^2}{\Delta r_1^2}}\right)\cr
\\ \nl &&=\frac{\left(P^2_{+}+P^2_{-}\right)^{1/4}}{(2^5 3)^{1/4}
(\pi^2 T_2 l_{11}^3 l\Lambda_1^{\psi})^{1/2}}\Delta r \hspace{.1cm}
\mbox{}_2F_{1}\left(1/2,1/2;3/2;\frac{\Delta r^2}{\Delta r_1^2}\right)
\\ \nl &&=\frac{\left(P^2_{+}+P^2_{-}\right)^{1/4}}{(2^5 3)^{1/4}
(\pi^2 T_2 l_{11}^3 l\Lambda_1^{\psi})^{1/2}}\Delta r_1 \arcsin\left(\frac{\Delta r}{\Delta r_1}\right).\ea
Obviously, it can be inverted to give
\ba\nl r_{scl}(\s) = 3l + (27/2)^{1/4}\frac{\left(P^2_{+}+P^2_{-}\right)^{1/4}}
{(\pi^2 T_2 l_{11}^3 l\Lambda_1^{\psi})^{1/2}} \sin\left[(8/3)^{1/2}\frac{\pi^2 T_2 l_{11}^3 l\Lambda_1^{\psi}}{\left(P^2_{+}+P^2_{-}\right)^{1/2}}\sigma\right].\ea

\newpage

\end{document}